\documentclass[twocolumn,preprintnumbers]{revtex4-2}

\usepackage{graphicx}
\usepackage{epsf}
\usepackage{color}
\usepackage{amsmath,amssymb}
\usepackage{ascmac}
\usepackage{bm}
\usepackage{color}
\usepackage{ulem}

\newcommand{\mix}{\rm{combi}}
\newcommand{\mixs}{c}

\newcommand{\cR}{{\cal R}}

\newcommand{\FRm}{{F_{c}}}

\newcommand{\FRs}{{F_{\bm 1}}}

\newcommand{\FRt}{{F_{\bar{\bm 3}}}}

\newcommand{\QQFRs}{{F^{\rm 2Q}_{\bm 1}}}

\newcommand{\QQQFRt}{{F^{\rm 3Q}_{\bar{\bm 3}}}}

\newcommand{\QQQQFRt}{{F^{\rm 4Q}_{\bar{\bm 3}}}}

\newcommand{\QQQQFRm}{{F^{\rm 4Q}_{c}}}

\newcommand{\rhosa}{\rho_{    {\bm 1}}^{\{1,3\}}}
\newcommand{\rhosb}{\rho_{    {\bm 1}}^{\{1,4\}}}
\newcommand{\rhos} {\rho_{    {\bm 1}}}
\newcommand{\rhoa} {\rho_{    {\bm 8}}}
\newcommand{\rhot} {\rho_{\bar{\bm 3}}}
\newcommand{\rhox} {\rho_{    {\bm 6}}}

\newcommand{\hrhos}{\hat{\rho}_{    {\bm 1}}}
\newcommand{\hrhoa}{\hat{\rho}_{    {\bm 8}}}
\newcommand{\hrhot}{\hat{\rho}_{\bar{\bm 3}}}
\newcommand{\hrhox}{\hat{\rho}_{    {\bm 6}}}

\newcommand{\hrhoai}{\hat{\rho}_{    {\bm 8}_i}}
\newcommand{\hrhoti}{\hat{\rho}_{\bar{\bm 3}_i}}
\newcommand{\hrhoxi}{\hat{\rho}_{    {\bm 6}_i}}

\newcommand{\dTQ}{d}
\newcommand{\hTQ}{h}
\newcommand{\dFQ}{d}
\newcommand{\hFQ}{h}

\newcommand{\Nadj}{{N_c^2-1}}

\usepackage{version}	
\begin{document}

\title{Lattice QCD study of color correlations between quarks in static multiquark systems}

\author{Toru T. Takahashi}
\affiliation{National Institute of Technology, Gunma College, Maebashi, Gunma
371-8530, Japan}
\author{Yoshiko Kanada-En'yo}
\affiliation{Department of Physics, Kyoto University, 
Sakyo, Kyoto 606-8502, Japan}

\date{\today}

 \begin{abstract}
  We study the color correlation between
  two static quarks in 3Q ($QQQ$) and 4Q ($QQ\bar Q\bar Q$) 
  multiquark systems at $T=0$ based on the 
  reduced two-body density matrices $\rho$ in color space.
  We perform quenched lattice QCD calculations with the Coulomb gauge
  adopting the standard Wilson gauge action,
  and the spatial volume is $L^3 = 32^3$ at $\beta = 5.8$,
  which corresponds to the lattice spacing $a=0.14$ fm 
  and the system volume $L^3=4.5^3$ fm$^3$.
  We evaluate the two-body color density matrix $\rho$ of static quarks,
  and investigate the dependence of color correlations on the quarks' spatial configuration.
  As a result, 
  we find that the color correlations depend on the minimal path length along a flux tube
  which connects two quarks under consideration.
  The color correlation between quarks quenches because of color leak 
  into the gluon field (flux tube) and finally approaches 
  the random color configuration in the large distance limit.
  We also find a ``universality'' in the flux-tube path length dependence of the color correlations
  for 2Q, 3Q, and 4Q ground-state systems.
Our results show that the color correlations of end-point quarks can be a clue to clarify
the internal structures of hadrons, including exotic (multiquark) hadrons.
 \end{abstract}

\maketitle

\section{Introduction}
\label{Sec.Introduction}

QuantumChromoDynamics (QCD) is the fundamental theory of the strong interaction,
and quarks are all confined in a totally color-singlet hadronic cluster
due to the color confinement phenomenon caused by QCD~\cite{Greensite:2011}.
There have been a lot of attempts to clarify the color confining nature of QCD,
and its nonperturbative dynamics is still attracting great interest.
In color singlet hadronic clusters, quarks' color flows into in-between gluon field
forming a confining flux tube among quarks~\cite{Bali:1994de,Bornyakov:2004uv,Cardoso:2011fq,Tiktopoulos:1976sj,Greensite:2001nx,Takahashi:2002bw,Okiharu:2004ve,Okiharu:2004wy}.
The in-between flux tube is a colored gluonic object created by end-point color sources, quarks,
and total systems are kept color singlet by quarks and gluons~\cite{Tiktopoulos:1976sj,Greensite:2001nx}.

In Refs.~\cite{Takahashi:2019ghj,Takahashi:2020bje,Takahashi:2024vff}, 
we studied color correlations in two static quark systems ($Q\bar Q$),
and found that the color charge initially associated with
quarks flows into the interquark region, and forms a confining flux tube as the total system size is enlarged.
This color transfer to the flux tube
can be regarded as a color charge leak from quarks to the gluon fields,
and is quantified as the color screening effect among quarks.
When quarks are located at distances close to each other,
color from a quark is absorbed by other quarks without any loss,
and color leak into gluon fields (flux tube) hardly occurs.
In this case, the color correlation among quarks is maximal.
As the system size is enlarged,
a physical gluonic flux tube grows and quarks' color is screened inside the tube,
which in turn weakens the quarks' color correlation.
This correlation quench is expressed as a mixture of a random color contribution
into quarks' color configuration,
in which {\it all the color combinations of the quark pair equally contribute}~\cite{Takahashi:2019ghj,Takahashi:2020bje,Takahashi:2024vff}.
At the large system size limit, 
the correlation disappears and the quarks' color configuration approaches
a random color configuration~\cite{Takahashi:2019ghj,Takahashi:2020bje,Takahashi:2024vff}.
This color transfer mechanism, from quarks to in-between gluons, is also found in 
a $Q\bar Q$ system accompanied by gluonic excitations~\cite{Takahashi:2024vff}.

While we have so far focused on systems consisting of static quark and antiquark ($Q\bar Q$),
quarks' color correlations in multiquark systems are also of great interest.
In multiquark systems, there can appear flux tubes that have ``junctions''~\cite{Bornyakov:2004uv,Cardoso:2011fq,Takahashi:2002bw,Okiharu:2004ve,Okiharu:2004wy},
and the color structure and the color flow inside hadronic clusters
would be more complicated than simple $Q\bar Q$ cases.

Several nontrivial issues arise for such multiquark systems.
In the $Q\bar Q$ case, the end-point quarks' color correlation
is quenched due to the color screening effect inside flux tubes,
and damps exponentially as a function of the gluon flux tube length $L$
(interquark distance $R$).
A question that arises here is
{\it how the quark pair color correlations in multiquark systems 
are affected by surrounding quarks and gluons.}
Does the color screening occur also in such systems?
If that is the case, {\it is the screening behavior described
based on the flux tube picture?}
Moreover, 4Q systems raise an intriguing question
about {\it how the color structures change across the flip-flop process,
the rearrangement of flux tubes.}
To answer these questions is
challenging and may give a clue to clarify
internal structures of exotic hadrons including excited hadrons and heavy quark hadrons.
To tackle these problems, we extend our analysis for $Q\bar Q$ systems
to multiquark systems.

In this paper, we define the two-body reduced color density matrix $\rho$ 
for 3Q ($QQQ$) and 4Q ($QQ\bar Q\bar Q$) systems,
and investigate the color structure of the multiquark systems.
According to the ansatz for the reduced density matrix $\rho$
proposed in Ref.~\cite{Takahashi:2019ghj},
we clarify the color correlation of two quarks inside the multiquark systems.

In Sec.~\ref{Sec.Formalism}, we give the formalism to compute
the reduced density matrix $\rho$ of multiquark systems.
The details of numerical calculations and ansatz for $\rho$
are also shown in Sec.~\ref{Sec.Formalism}.
Results are presented in Sec.~\ref{Sec.Results3Q}, Sec.~\ref{Sec.Results4Q}, and Sec.~\ref{Sec.ResultsEE}.
Sec.~\ref{Sec.Summary} is devoted to the summary and concluding remarks.

\section{Formalism}
\label{Sec.Formalism}

\subsection{reduced 2-body density matrix and quarks' color correlation}

We investigate the color correlation between two static (anti)quarks
via the two-body density matrix $\rho$ evaluated in terms of quarks' color configuration.
A density matrix $\rho$ defined in such a way corresponds to the reduced density matrix obtained by integrating out the other degrees of freedom (d.o.f.) ({\it e.g.,} gluonic d.o.f.)
in the full density matrix.

In the following, we exemplify the formalism using quarks $Q_i$.
When a system containing antiquarks is considered,
readers should substitute $Q_i$ with $\bar Q_i$ as appropriate.

For an $N$-quark ($N$Q) system, we choose two quarks of them
in order to investigate the color correlation of quark pairs in $N$Q systems.
A full density operator $\hat\rho_{\rm full}({\cR})$ for an $N$Q system
is defined as
\begin{equation}
\hat\rho_{\rm full}({\cR}) = |NQ({\cR})\rangle \langle NQ({\cR})|.
\end{equation}
Here $|NQ({\cR})\rangle$ represents a quantum state of an $N$Q system.
${\cR}$ is a vector set defined as ${\cR}\equiv\{\vec{r_1},\vec{r_2},\vec{r_3},...\}$, where
the vector $\vec{r_i}$ 
denotes the $i$-th quark's ($Q_i$'s) spatial position.
When we measure the color correlation between the $i$- and $j$-th quarks, 
$Q_k$ ($k\neq i,j$) are all ``spectator'' quarks 
in the sense that we do not see their colors, and that their dynamics are traced out.
Gluon fields' color is not seen either, and is also traced out in the lattice QCD simulation.
Then, the quantum state we will find for the target quark pair ($Q_iQ_j$ pair) is expressed by the reduced density operator
\begin{equation}
\hat\rho({\cR}) = \int {\cal D}Q'{\cal D}G\ \langle Q'G|\hat\rho_{\rm full}({\cR})| Q'G\rangle.
\end{equation}
Here, $Q'$
represents the d.o.f. for all the spectator quarks other than $Q_i$ and $Q_j$ ($|Q'\rangle \equiv \Pi_{k\neq i,j}|Q_k(\vec r_k)\rangle$),
and $G$ denotes all the gluonic d.o.f.
Note that the reduced density operator $\hat\rho({\cR})$ still depends on the spatial positions of spectator quarks, $\vec r_k$ ($k\neq i,j$).
The matrix elements $\rho({\cR})_{ab,cd}$ of the reduced density operator,
where $a$ and $c$ ($b$ and $d$) are quark's color indices,
are expressed as
\begin{equation}
\rho({\cR})_{ab,cd} = \langle Q_i^a(\vec{r_i}) Q_j^b(\vec{r_j})|\hat\rho({\cR})|Q_i^c(\vec{r_i}) Q_j^d(\vec{r_j}) \rangle.
\end{equation}
$|Q_i^c(\vec{r_i}) Q_j^d(\vec{r_j}) \rangle$ represents a state
in which the $i$- and $j$-th quarks located at $\vec r_i$ and $\vec r_j$ have color indices $c$ and $d$, respectively.
In this case, $k$-th quarks ($k\neq i,j$) are all spectators.
In $\rho({\cR})_{ab,cd}$, we omit the quarks' indices, $i$ and $j$, as long as there is no misunderstanding.

Then, $\rho({\cR})$ is an $N_c\times N_c$ square matrix.
The density matrix $\rho({\cR})$ is evaluated using only target quarks' color d.o.f., and
does not explicitly contain spectator quarks' and gluon's color d.o.f. in this construction:
the spectator quarks' and gluonic d.o.f. are ``integrated out'' in the lattice calculation,
and $\rho({\cR})$ can be regarded as a reduced density matrix~\cite{Takahashi:2019ghj}.

In this prescription, we directly measure the quarks' color, hence we have to fix the gauge. In our present study, we employ the Coulomb gauge,
which is closely related to the confinement physics
~\cite{Reinhardt:2017pyr}.

\subsection{Ansatz for reduced density matrix $\rho_{ab,cd}({\cR})$}

In the series of our papers~\cite{Takahashi:2019ghj,Takahashi:2020bje,Takahashi:2024vff},
we investigated the color structure of quark and antiquark
in static $Q\bar Q$ systems with and without gluonic excitations.
As a result, we found that 
the color correlation is randomized as the interquark distance $R$ increases for both systems.
In a ground-state $Q\bar Q$ system with no gluonic excitation,
the color configuration of a $Q\bar Q$ pair
can be well represented by the density operator~\cite{Takahashi:2019ghj,Takahashi:2020bje}
\begin{eqnarray}
\hat\rho^{\rm ansatz}(R)
=
F_1(R)\hat\rho_{\bm 1} + (1-F_1(R))\hat\rho^{\rm rand},
\end{eqnarray}
where $R$ is the interquark distance.
Here, $\hat\rho_{\bm 1}=|{\bm 1}\rangle\langle{\bm 1}|$ is a density operator for a $Q\bar Q$ pair 
that forms a purely color singlet configuration in the Coulomb gauge,
and $\hat\rho^{\rm rand}$ is that for a random color state
where one singlet state and eight octet states enter with equal weights.
The ``initial'' color configuration of a $Q\bar Q$ pair at $R\sim 0$
was found to be color singlet, and it leads to $F_1(R)\sim 1$.
On the other hand, when the $Q\bar Q$ distance $R$ is large,
the color configuration is randomized and $F_1(R)$ approaches zero.

This ansatz with $\hat\rho^{\rm rand}$
is also applicable for $Q\bar Q$ systems with gluonic excitations ($Q\bar QG$ systems).
By replacing the initial color configuration $\hrhos$
with $\hrhoa = \frac18 \sum_i |{\bm 8}_i\rangle\langle{\bm 8}_i|$
representing the color octet state,
which ``initially'' dominates the $Q\bar QG$ system at $R\sim 0$,
the color configuration of a $Q\bar Q$ pair in a $Q\bar QG$ system
can be expressed by the density operator~\cite{Takahashi:2024vff}
\begin{eqnarray}
\hat\rho^{\rm ansatz}(R)
=
F_8(R)\hat\rho_{\bm 8} + (1-F_8(R))\hat\rho^{\rm rand}.
\end{eqnarray}

From these observations, 
$F_1(R)$ and $F_8(R)$ in the ansatz are considered to be a {\it residual rate of the maximally correlated configuration} of a (hybrid) $Q\bar Q$ pair,
which is expected when no gluon flux tube is present at $R\rightarrow 0$.
In fact, at $R\rightarrow 0$, 
the color flowing out of one quark is fully absorbed by the antiquark with no color leak,
and the $Q\bar Q$ color correlation is maximized.
As a physical flux tube grows between quarks, the color from quarks is absorbed also by in-between gluons,
and the color correlation between end-point quarks is weakened.
The exponential decay of the ``residual rate'' $F_1(R)$ and $F_8(R)$ implies
quarks' color screening by in-between gluons,
and it was investigated in detail in the previous papers~\cite{Takahashi:2019ghj,Takahashi:2020bje,Takahashi:2024vff}.

By analogy with the previous results of $Q\bar Q$ systems, we consider the case of multiquark systems.
When all the quarks are located in close to each other and there is no color leak to a gluon flux tube,
any two quarks in multiquark systems are expected to form 
``maximally correlated'' (MC) color states depending on a system under investigation,
and the random configuration would enter as in-between gluon flux tubes grow.
Note that this MC state is nothing but a quarks' color configuration that appears
when no color leak from quarks to gluon fields occurs at ${\cR}\rightarrow 0$ ($|r_i|\rightarrow 0$ for all $i$).
With the density operator $\hat\rho^{\rm MC}$ for this MC state,
we introduce the color density operator $\hat\rho^{\rm ansatz}$ for a quark pair as
\begin{eqnarray}
\hat\rho^{\rm ansatz}({\cR})
=
F_{\rm MC}({\cR})\hat\rho^{\rm MC} + (1-F_{\rm MC}({\cR}))\hat\rho^{\rm rand}.
\label{Eq.GAnsatz}
\end{eqnarray}
Here, $F_{\rm MC}({\cR})$ represents the ``residual rate'' of the MC state expected when there is no color leak to gluon flux tubes at ${\cR}\rightarrow 0$.
The color configuration of a quark pair for any ${\cR}$ can be solely represented by $F_{\rm MC}({\cR})$ in this ansatz.

\subsubsection{$QQ$ color correlation in 3Q and 4Q systems}

First, we consider the case when we measure the color correlation
between two quarks ($QQ$) in multiquark (3Q and 4Q) systems.
A color configuration of any $QQ$ pair is decomposed
into color antitriplet and sextet states.
Let $\hrhoti$ and $\hrhoxi$ be the density operators
for color antitriplet states $|{\bar{\bm 3}_i}\rangle$
and sextet states $|{\bar{\bm 6}_i}\rangle$ in the Coulomb gauge
respectively expressed as
\begin{eqnarray}
\hrhoti &=& |{\bar{\bm 3}_i}\rangle \langle {\bar{\bm 3}_i}| \ \   (i=1\sim 3),
\\
\hat\rho_{{\bm 6}_i} &=& |{\bm 6}_i\rangle \langle {\bm 6}_i|
\ \ (i =1\sim 6).
\end{eqnarray}
These density operators can be expressed
in a matrix form
that diagonalizes $\hrhoti$ and $\hrhoxi$.
For example, for $i=2$,
\begin{equation}
 \hrhoti
 =
 {\rm diag}(0,1,0,...,0)_{\bm \alpha}.
\end{equation}
Here, the subscript ``${\bm \alpha}$'' means that the matrix is expressed
in terms of $QQ$'s color representation
with the vector set of ${\bm \alpha}=\{|{\bar{\bm 3}_1}\rangle,...,|{\bm 6_1}\rangle,...|{\bm 6_6}\rangle\}$.

Taking into account the color SU(3) symmetry,
it is convenient to define the averaged density operators $\hrhot$ and $\hrhox$ for 
antitriplet and sextet states as
\begin{eqnarray}
\hrhot = \frac{1}{3}\sum_{i=1}^{3}\hrhoti,
\ \ \ 
\hrhox = \frac{1}{6}\sum_{i=1}^{6}\hrhoxi.
\end{eqnarray}

As the interquark distance increases, an uncorrelated state,
{\it i.e.,} a randomized-color state,
mixes in $\rho({\cR})$ due to the color screening effect by in-between gluons.
Such a random-color state contains
all the $N_c^2$ components with equal weights,
and its density operator is given as 
\begin{eqnarray}
 \hat\rho^{\rm rand}
&=&
 \frac{1}{N_c^2}\left(
\sum_{i=1}^{3}\hrhoti
 +
\sum_{i=1}^{6}\hrhoxi\right)
\\
&=&
\frac{1}{N_c^2}{\hat I}
 =
{\rm diag}\left(\frac{1}{N_c^2},\frac{1}{N_c^2},...,\frac{1}{N_c^2}\right)_{\bm \alpha}.
\end{eqnarray}

We are mainly interested in the ground-state 3Q and 4Q systems, and
their color structures 
will be $([Q_1Q_2]Q_3)$ and $([Q_1Q_2][\bar Q_3\bar Q_4])$
when all the quarks are close to each other at ${\cR}\rightarrow 0$.
Here, the square and curly brackets respectively show that 
inside quarks form a color (anti)triplet and singlet configuration.
In this case, any $QQ$ pair is considered
to form a color-antitriplet state ($|{\bar{\bm 3}}_i\rangle$)
corresponding to the MC configuration in both systems,
hence we choose $\hrhot$ for $\hat\rho^{\rm MC}$ in Eq.~(\ref{Eq.GAnsatz})
to measure $QQ$ correlations in 3Q and 4Q systems.
Letting the residual rate of the MC color antitriplet state be $\FRt(\cR)$
and the fraction of the random contribution be $(1-\FRt(\cR))$,
the density operator for a spatial configuration ${\cR}$ in this ansatz is written as
\begin{eqnarray}
 \hat\rho_{0,{\bar{\bm 3}}}^{\rm ansatz}({\cR})
  &=&
  \FRt(\cR)\hrhot+(1-\FRt(\cR))\hat\rho^{\rm rand}.
  \label{Eq.TAnsatz}
\end{eqnarray}
Here, the subscript $(0,{\bar{\bm 3}})$ means that
the density operator $\hat\rho_{0,{\bar{\bm 3}}}^{\rm ansatz}({\cR})$ is dominated by the color antitriplet contribution $\hrhot$
at ${\cR}\rightarrow 0$.
The matrix form $\rho({\cR})_{\bm \alpha}$ of $\hat\rho_{0,{\bar{\bm 3}}}^{\rm ansatz}({\cR})$ is expressed as
\begin{widetext}
\begin{eqnarray}
  \hat\rho_{0,{\bar{\bm 3}}}^{\rm ansatz}(R)
  &=&
  \FRt(\cR)\hrhot+(1-\FRt(\cR))\hat\rho^{\rm rand}
  \label{Eq.TAnsatzs01}
  \\
  &=&
  \left(\FRt(\cR)+\frac{3}{N_c^2}(1-\FRt(\cR))\right) \hrhot
  +
  \left(\frac{6}{N_c^2}(1-\FRt(\cR))\right) \hrhox
  \label{Eq.TAnsatzs02}
  \\
  &=&
  {\rm diag}\left(
	     \frac13\FRt(\cR)+\frac{1}{N_c^2}(1-\FRt(\cR)),...,\frac{1}{N_c^2}(1-\FRt(\cR))
	     ,...
 	    \right)_{\bm \alpha}
  \label{Eq.TAnsatzs03}
  \\
  &\equiv&
  {\rm diag}\left(
	     \rho({\cR})_{\bar{\bm 3}_1,\bar{\bm 3}_1},...,
	     \rho({\cR})_{    {\bm 6}_1,    {\bm 6}_1}
	     ,...
	    \right)_{\bm \alpha}.
\end{eqnarray}
\end{widetext}
In this ansatz,
\begin{eqnarray}
 \begin{cases}
 \rho({\cR})_{\bar{\bm 3}_1,\bar{\bm 3}_1}
 =
 \rho({\cR})_{\bar{\bm 3}_2,\bar{\bm 3}_2}
 =
 \rho({\cR})_{\bar{\bm 3}_3,\bar{\bm 3}_3}
\\
 \rho({\cR})_{{\bm 6}_1,{\bm 6}_1}
 =
 \rho({\cR})_{{\bm 6}_2,{\bm 6}_2}
 =...=
 \rho({\cR})_{{\bm 6}_6,{\bm 6}_6}
\\  
 \rho({\cR})_{{\bm a},{\bm b}}=0\ \ ({\rm for}\ {\bm a}\neq{\bm b})
 \end{cases}
\label{Eq.conditions36}
\end{eqnarray}
and the normalization condition
\begin{eqnarray}
{\rm Tr}\  \rho({\cR})
=
 \rhot({\cR})
+
 \rhox({\cR})
=1
\end{eqnarray}
are satisfied at any ${\cR}$.
Here,
$\rhot({\cR})$ and $\rhox({\cR})$ are the summations of 
the diagonal matrix elements,
\begin{eqnarray}
\rhot({\cR})\equiv \sum_{i=1}^{3}\rho({\cR})_{\bar{\bm 3}_i,\bar{\bm 3}_i},\ \ \  \rhox({\cR})\equiv \sum_{i=1}^{6}\rho({\cR})_{{\bm 6_i},{\bm 6_i}},
\end{eqnarray}
which indicate the antitriplet and sextet components.
We directly compute $\rhot$ and $\rhox$ using lattice QCD calculations, and
the residual rate $\FRt(\cR)$ can be extracted from $\rhot({\cR})$ or $\rhox({\cR})$ as
\begin{eqnarray}
\FRt(\cR) =  1-\frac{N_c^2}{6}\rhox({\cR}).
\end{eqnarray}

Indeed, it was found that this ansatz reproduces the density matrix element 
for the ground-state $Q\bar Q$ system
evaluated by lattice QCD calculation with a very good accuracy~\cite{Takahashi:2019ghj}.

\subsubsection{$Q\bar Q$ color correlation in 4Q systems}

The color configuration of a $Q\bar Q$ pair in 4Q ($QQ\bar Q\bar Q$) systems is classified into 
singlet and octet color configurations, 
$|{\bm 1}\rangle$ and $|{\bm 8}_i\rangle$.
Then we define the density operators
for a pair forming a color singlet state $|{\bm 1}\rangle$ 
and an octet state $|{\bm 8_i}\rangle$ in the Coulomb gauge
as
\begin{equation}
\hat\rhos = |{\bm 1}\rangle \langle {\bm 1}|,
\ \ \ 
\hat\rho_{{\bm 8}_i} = |{\bm 8}_i\rangle \langle {\bm 8}_i|
\ \ (i =1\sim \Nadj).
\end{equation}

As before, we define the averaged density operator $\hrhoa$
as
\begin{eqnarray}
\hrhoa = \frac{1}{\Nadj}\sum_{i=1}^{\Nadj}\hrhoai.
\end{eqnarray}

In the case when we measure the color correlation inside a $Q\bar Q$ pair
in a 4Q system,
the ``initial'' MC color configuration of a $Q\bar Q$ pair
would be neither $|{\bm 1}\rangle$ nor $|{\bm 8}_i\rangle$,
but a combination of them.
Especially when four quarks form a ``genuine'' multiquark state
$\left([Q_1Q_2][\bar Q_3 \bar Q_4]\right)$,
$\hat\rho^{\rm MC}$ in Eq.(\ref{Eq.GAnsatz}) would be~\cite{Okiharu:2004ve}
\begin{equation}
\hat\rho^{\rm MC} = \frac13 \hrhos + \frac23 \hrhoa \equiv \hat\rho^{\mix},
\label{Eq.rhomix}
\end{equation}
leading to the ansatz for 4Q systems
\begin{eqnarray}
 \hat\rho_{0,\mixs}^{\rm ansatz}({\cR})
  &=&
  \FRm(\cR)\hat\rho^{\mix}
  +
  (1-\FRm(\cR))\hat\rho^{\rm rand}
\label{Eq.MAnsatz}
  \\
  &=&
  \FRm(\cR)\left(\frac13 \hrhos + \frac23 \hrhoa.\right)
  \nonumber
  \\
  &+&
  (1-\FRm(\cR))\hat\rho^{\rm rand}.
\end{eqnarray}
Here, the subscript $(0,\mixs)$ means that
the density operator $\hat\rho_{0,\mixs}^{\rm ansatz}({\cR})$ 
is expressed by the residual rate $\FRm(\cR)$ of the MC color configuration $\hat \rho^{\mix}$,
and $\hat \rho^{\mix}$ dominates $\hat\rho_{0,\mixs}^{\rm ansatz}({\cR})$ 
at ${\cR}\rightarrow 0$ corresponding to $\FRm(\cR) \sim 1$.
In the matrix form ${\bm \beta}$ expressed
with the vector set of ${\bm \beta}=\{|{\bm 1}\rangle,|{\bm 8_1}\rangle,...|{\bm 8_8}\rangle\}$,
$\hat\rho_{0,\mixs}^{\rm ansatz}({\cR})$ can be explicitly expressed as
\begin{widetext}
\begin{eqnarray}
  \hat\rho_{0,\mixs}^{\rm ansatz}({\cR})
  &=&
  \FRm(\cR)\left(\frac13 \hrhos + \frac23 \hrhoa \right)+(1-\FRm(\cR))\hat\rho^{\rm rand}
  \label{Eq.MAnsatzs01}
  \\
  &=&
  \left(\frac13\FRm(\cR) +\frac{1}{N_c^2}(1-\FRm(\cR))\right) \hat\rhos
  +
  \left(\frac23\FRm(\cR) +\frac{\Nadj}{N_c^2}(1-\FRm(\cR))\right) \hrhoa
  \label{Eq.MAnsatzs02}
  \\
  &=&
  {\rm diag}\left(
	     \frac13\FRm(\cR) +\frac{1}{N_c^2}(1-\FRm(\cR))
            ,\frac{2}{3(\Nadj)}\FRm(\cR) +\frac{1}{N_c^2}(1-\FRm(\cR))
	    ,...
 	    \right)_{\bm \beta}
  \label{Eq.MAnsatzs03}
  \\
  &\equiv&
  {\rm diag}\left(
	     \rho(R)_{{\bm 1},{\bm 1}},
	     \rho(R)_{{\bm 8}_1,{\bm 8}_1}
	     ,...
	    \right)_{\bm \beta}.
\end{eqnarray}
\end{widetext}

Again,
\begin{eqnarray}
 \begin{cases}
 \rho({\cR})_{{\bm 8}_1,{\bm 8}_1}
 =
 \rho({\cR})_{{\bm 8}_2,{\bm 8}_2}
 =...=
 \rho({\cR})_{{\bm 8}_\Nadj,{\bm 8}_\Nadj}
\\  
 \rho({\cR})_{{\bm a},{\bm b}}=0\ \ ({\rm for}\ {\bm a}\neq{\bm b})
 \end{cases}
\label{Eq.conditions18}
\end{eqnarray}
and the normalization condition
\begin{eqnarray}
{\rm Tr}\  \rho({\cR})
=
 \rhos({\cR})
+
 \rhoa({\cR})
=1,
\end{eqnarray}
\begin{eqnarray}
\rhos({\cR})\equiv \rho({\cR})_{{\bm 1},{\bm 1}},\ \ \ 
\rhoa({\cR})\equiv \sum_{i=1}^{\Nadj}\rho({\cR})_{{\bm 8_i},{\bm 8_i}}.
\end{eqnarray}
are satisfied at any ${\cR}$.

$\FRm(\cR)$ can be reconstructed from $\rhos({\cR})$ or $\rhoa({\cR})$ as
\begin{eqnarray}
\FRm(\cR) =  \frac{3}{N_c^2-3}\left(N_c^2\rhos({\cR}) - 1\right).
\end{eqnarray}

\subsection{Lattice QCD formalism}

\begin{figure}[h]
 \includegraphics[width=7cm]{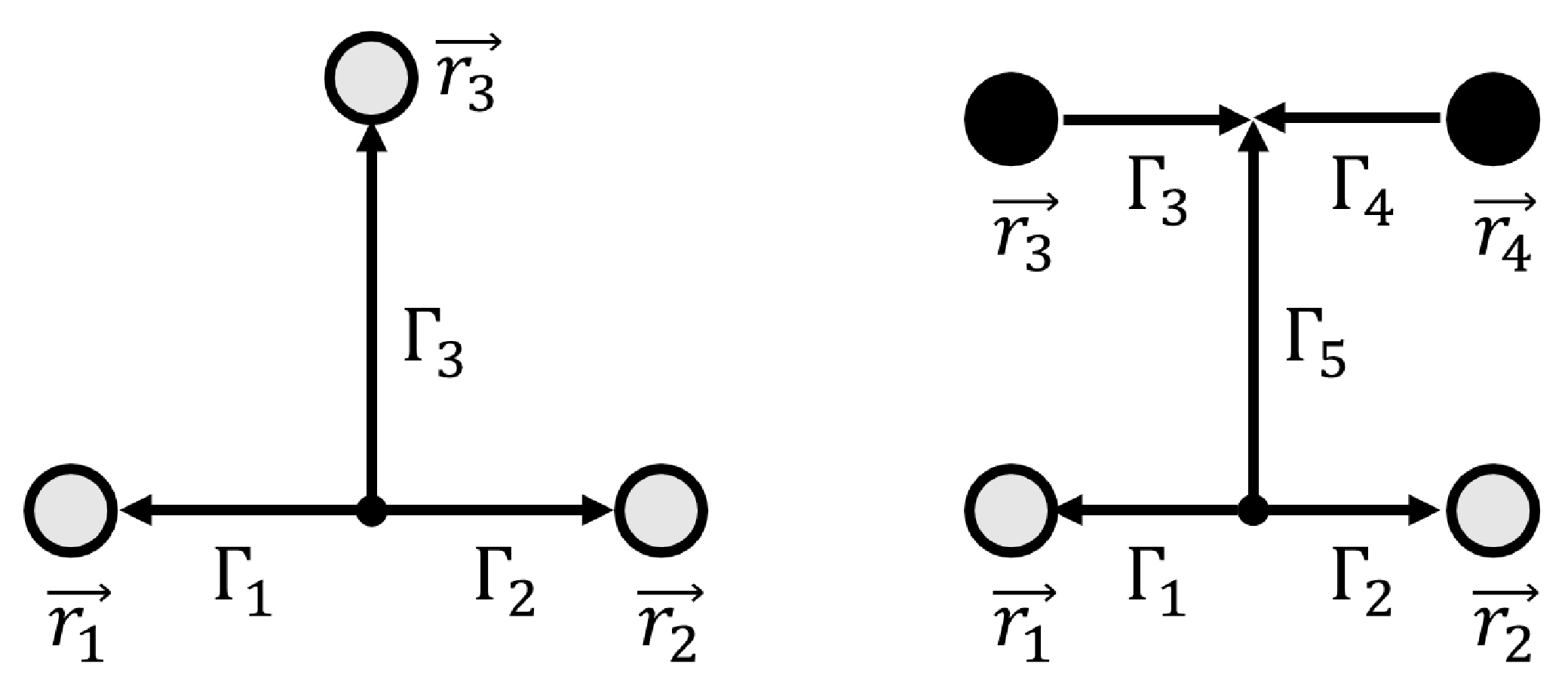}
\caption{\label{Fig.OP3Q4Q}
The paths $\Gamma_i$ that define the path-ordered products $U(\Gamma_i,t)\in {\rm SU(3)}$ are shown
for 3Q and 4Q cases.
}
\end{figure}
 
For the 3Q case, we locate static three quarks at
${\cR}=\{\vec r_1, \vec r_2, \vec r_3\}$ in a $xy$ plane,
and define paths that stem from a junction point to $\vec r_i$ as $\Gamma_i$ ($i=1,2,3$)
as depicted in the left panel in Fig.~\ref{Fig.OP3Q4Q}.
With the path-ordered products $U(\Gamma_i,t)\in {\rm SU(3)}$ at the Euclidean time $t$ defined as
\begin{equation}
U(\Gamma_i,t)\equiv P\exp\left\{ig\int_{\Gamma_i} A_\mu dx_\mu\right\},
\end{equation}
we define 3Q operator $Q^{\rm 3Q}_{c_1c_2c_3}({\cR},t)$ as
\begin{equation}
Q^{\rm 3Q}_{c_1c_2c_3}({\cR},t)\equiv
\varepsilon^{abc}
U_{c_1a}(\Gamma_1,t)
U_{c_2b}(\Gamma_2,t)
U_{c_3c}(\Gamma_3,t),
\end{equation}
\begin{equation}
Q^{\rm 3Q}_{c_1c_2c_3}({\cR},t)^\dagger\equiv
\varepsilon^{abc}
U^\dagger_{ac_1}(\Gamma_1,t)
U^\dagger_{bc_2}(\Gamma_2,t)
U^\dagger_{cc_3}(\Gamma_3,t).
\end{equation}
Here, the subscripts denote the color indices,
and subscripts that appear twice are to be contracted.
We can construct the correlator $C^{\rm 3Q}(T,{\cR})$ for a static 3Q system as
\begin{eqnarray}
C^{\rm 3Q}({\cR},T)=Q^{{3Q}}_{c_1'c_2'c_3'}({\cR},T)^\dagger
Q^{\rm 3Q}_{c_1c_2c_3}({\cR},0)
\nonumber \\
\times
U_4(\vec r_1,T)_{c_1'c_1}
U_4(\vec r_2,T)_{c_2'c_2}
U_4(\vec r_3,T)_{c_3'c_3}.
\end{eqnarray}
$U_4(\vec r,T)$ is the path-ordered protuct in the temporal direction
explicitly written as
\begin{equation}
U_4(\vec r,T) = P\exp\left\{ig\int_{(\vec r,0)}^{(\vec r,T)} A_4 dx_4 \right\}.
\end{equation}
When we measure the color of $Q_i$,
we replace the temporal part, $U_4(\vec r_i,T)$, in $C^{3Q}({\cR},T)$ with
\begin{eqnarray}
\tilde U_4(\vec r_i;T)_{ab,c'c}
&\equiv&
P\exp\left\{ig\int_{(\vec r_i,0)}^{(\vec r_i,T/2)} A_4 dx_4 \right\}_{c'a}
\nonumber \\
&\times&
P\exp\left\{ig\int_{(\vec r_i,T/2)}^{(\vec r_i,T)} A_4 dx_4 \right\}_{bc},
\end{eqnarray}
in which $Q_i$'s color is measured at the temporal point $t=T/2$
by inserting a quark operator $\hat{Q}^{\dagger a}_i(\vec r_i) \hat{Q}^b_i(\vec r_i)$ at $t=T/2$ plane.
For example, when we measure the $Q_1Q_2$ color correlation,
we replace $U_4(\vec r_1,T)_{c_1'c_1}$ and $U_4(\vec r_2,T)_{c_2'c_2}$ in $C^{3Q}({\cR},t)$ with
$\tilde U_4(\vec r_1;T)_{ac,c_1'c_1}$
and
$\tilde U_4(\vec r_2;T)_{bd,c_2'c_2}$,
and define
\begin{eqnarray}
&&
\tilde C^{\rm 3Q}_{ab,cd}({\cR},T)
\equiv
Q^{{3Q}}_{c_1'c_2'c_3'}({\cR},T)^\dagger
Q^{\rm 3Q}_{c_1c_2c_3}({\cR},0)
\nonumber \\
&\times&
\tilde U_4(\vec r_1,T)_{ac,c_1'c_1}
\tilde U_4(\vec r_2,T)_{bd,c_2'c_2}
U_4(\vec r_3,T)_{c_3'c_3}.
\end{eqnarray}

For the 4Q case, we locate static quarks and antiquarks at
${\cR}=\{\vec r_1, \vec r_2, \vec r_3, \vec r_4 \}$ in a $xy$ plane,
and define paths that stem from a junction to $\vec r_i$ or to another junction as $\Gamma_i$ ($i=1\sim 5$)
as depicted in the right panel in Fig.~\ref{Fig.OP3Q4Q}.
With the path-ordered protucts $U(\Gamma_i,t)$,
we define 4Q operator $Q^{\rm 4Q}_{c_1c_2,c_3c_4}({\cR},t)$ as
\begin{eqnarray}
&&
Q^{\rm 4Q}_{c_1c_2,c_3c_4}({\cR},t)
\nonumber \\
&\equiv&
\varepsilon^{abc}
\varepsilon^{a'b'c'}
U_{c_1a}(\Gamma_1,t)
U_{c_2b}(\Gamma_2,t)
\nonumber \\
&\times&
U_{a'c_3}(\Gamma_3,t)
U_{b'c_4}(\Gamma_4,t)
U_{c'c}(\Gamma_5,t).
\end{eqnarray}
We construct the correlator $C^{\rm 4Q}(T,{\cR})$ for a static 4Q system as
\begin{eqnarray}
&&
C^{\rm 4Q}({\cR},T)=
Q^{{4Q}}_{c_1'c_2',c_3'c_4'}({\cR},T)^\dagger
Q^{\rm 4Q}_{c_1c_2,c_3c_4}({\cR},0)
\nonumber \\
&\times&
U_4(\vec r_1,T)_{c_1'c_1}
U_4(\vec r_2,T)_{c_2'c_2}
U_4(\vec r_3,T)^\dagger_{c_3'c_3}
U_4(\vec r_4,T)^\dagger_{c_4'c_4}.
\end{eqnarray}
As the 3Q case,
replacing two of the temporal link variables in $C^{4Q}({\cR},T)$ with
$\tilde U_4(\vec r_i;T)_{ab,c'c}$ or $\tilde U_4(\vec r_i;T)^\dagger_{ab,c'c}$,
we can measure the color structure of internal two (anti)quarks and construct $\tilde C^{\rm 4Q}_{ab,cd}({\cR},T)$,
from which we compute the two-body color density matrix $\rho({\cR})$ for 4Q systems.

When $\tilde C^{N{\rm Q}}_{ab,cd}({\cR},T)$ couples only to the ground state,
$\langle \tilde C^{N{\rm Q}}_{ab,cd}({\cR},T) \rangle$ is then expressed as
\begin{eqnarray}
&&
\langle \tilde C^{N{\rm Q}}_{ab,cd}({\cR},T) \rangle
\nonumber \\
&=&
A
\langle NQ|
e^{- \frac12\hat{H} T}
 \hat{Q}^{\dagger a}_i(\vec r_i) \hat{Q}^{\dagger b}_j(\vec r_j)
 \hat{Q}^c_i(\vec r_i) \hat{Q}^d_j(\vec r_j)
e^{- \frac12\hat{H} T}
| NQ \rangle
\nonumber \\
&=&
Ae^{- E_0 T}
\langle NQ|
 \hat{Q}^{\dagger a}_i(\vec r_i) \hat{Q}^{\dagger b}_j(\vec r_j)
 \hat{Q}^c_i(\vec r_i) \hat{Q}^d_j(\vec r_j)
| NQ\rangle
\nonumber \\
&=&
Be^{- E_0 T}
\int{\cal D}Q'{\cal D}G \times 
\nonumber \\
&&
\langle NQ|
Q^a_i(\vec r_i) Q^b_j(\vec r_j)Q'G\rangle
\langle Q^c_i(\vec r_i) Q^d_j(\vec r_j)Q'G
| NQ\rangle
\nonumber \\
&=&
Be^{- E_0 T}\rho({\cR})_{ab,cd},
\end{eqnarray}
where $E_0$ is the ground state energy for the $N$Q system.
$Q'$ and $G$ again denote spectator quarks' and gluon's d.o.f.
Normalizing $\langle \tilde C^{N{\rm Q}}_{ab,cd}({\cR},T) \rangle$ so that
${\rm Tr}\ \langle \tilde C^{N{\rm Q}}_{ab,cd}({\cR},T) \rangle =\sum_{ab} \langle \tilde C^{N{\rm Q}}_{ab,ab}({\cR},T) \rangle = 1$,
we finally obtain the two-body color density matrix $\rho({\cR})$ 
for the ground state $N$Q state
whose trace is unity (${\rm Tr}\ \rho({\cR})=1$).

\subsection{Lattice QCD parameters}

We perform quenched calculations of
reduced density matrices of static quark pairs in multiquark (3Q, 4Q) systems
adopting the standard Wilson gauge action.
The gauge configurations are generated
on the spatial volume of $V = 32^3$ with the gauge couplings $\beta = 5.8$,
which corresponds to $V = 4.5^3$ [fm$^3$]. The temporal extent is $N_t = 32$.
All the gauge configurations are gauge-fixed with the Coulomb gauge condition.
While finite volume effects would still remain for $V = 4.5^3$ [fm$^3$]~\cite{Takahashi:2019ghj}, 
a detailed study taking care of such systematic errors 
is beyond the scope of the present paper, 
in which the color structure of multiquark systems is clarified for the first time.

\section{Lattice QCD results (3Q)}
\label{Sec.Results3Q}

\subsection{spatial configuration}

\begin{figure}[h]
 \includegraphics[width=2.5cm]{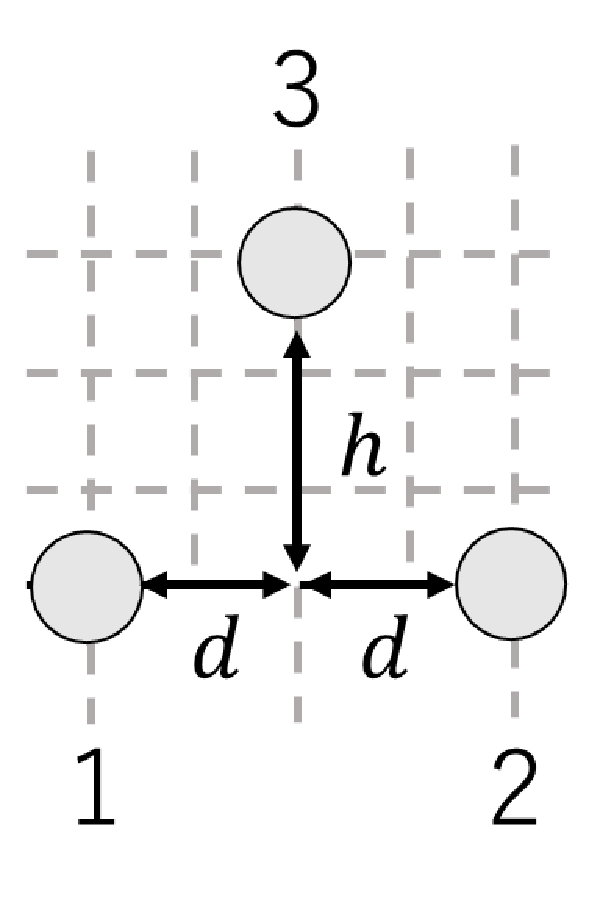}
\caption{\label{Fig.SPconf3Q}
Three quarks are located at $(+\dTQ,0,0)$, $(-\dTQ,0,0)$, $(0,+\hTQ,0)$
with integers $\dTQ$ and $\hTQ$ on the lattice.
When the correlation between 1- and 2-quarks (1- and 3-quarks) is measured, it is on-axis (off-axis) correlation.
}
\end{figure}

We locate three quarks at $(+\dTQ,0,0)$, $(-\dTQ,0,0)$, $(0,+\hTQ,0)$ with integers $\dTQ$ and $\hTQ$
in a $xy$ plane on the lattice as seen in Fig.~\ref{Fig.SPconf3Q}.
When the color correlation between $Q_1$ and $Q_2$ ($Q_1$ and $Q_3$) is measured, 
it is an on-axis (off-axis) correlation.
We adopt $1\leqq \dTQ \leqq 4$ and $1\leqq \hTQ \leqq 8$ for sets of ($\dTQ$,$\hTQ$),
totally 32 different spatial configurations.

\subsection{Functional form of the ansatz}

In the previous studies of $Q\bar Q$ systems, all the physical quantities were discussed
based on the interquark distance $R$.
It was actually found that the ``initial'' quarks' color is screened by the in-between gluon flux tube,
then the interquark distance $R$, which just coincides with the in-between flux tube length, is suitable for the analysis.
In the analysis of 3Q systems,
taking into account that color screening or color absorption occurs inside a flux tube,
we evaluate physical quantities based on the length $L$ of the flux-tube path between two quarks
of which we investigate the color structure.
As for the flux-tube shape, 
we consider the Y-type confining flux tube appearing in the ground-state 3Q systems~\cite{Bornyakov:2004uv,Takahashi:2002bw}.

\begin{figure}[h]
 \includegraphics[width=7cm]{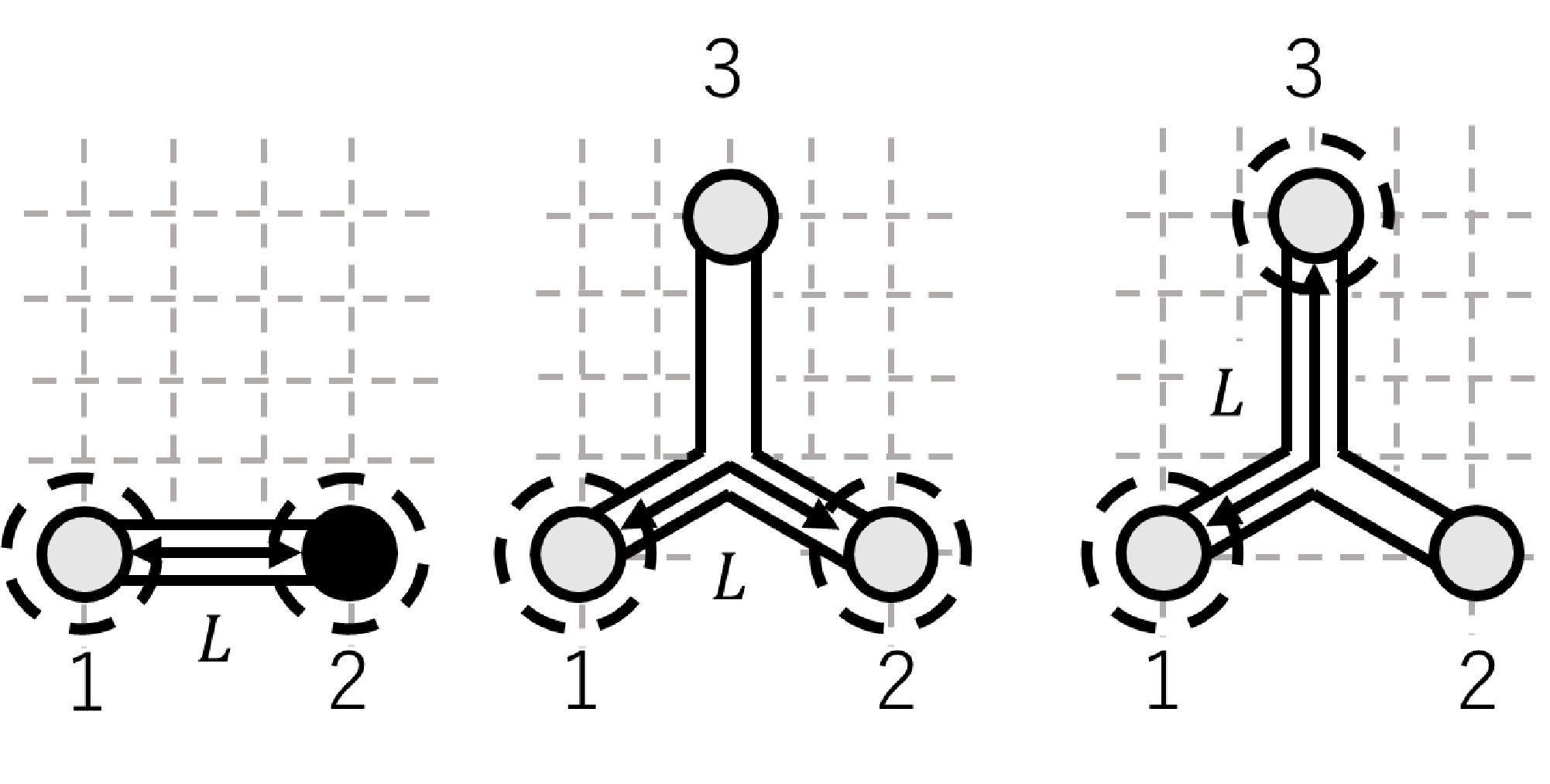}
\caption{\label{Fig.Fform01}
Schematic pictures which express the definition of $L$ in $Q\bar Q$ and 3Q systems.
Two (anti)quarks of which we measure the color correlation are enclosed in dotted circles.
}
\end{figure}

Figure~\ref{Fig.Fform01} shows schematic pictures of $L$.
Two (anti)quarks of which we measure the color correlation are enclosed in dotted circles.
In the $Q\bar Q$ case (left panel), the flux-tube path length $L$ coincides with the interquark distance $R$.
In 3Q systems (middle and right panels), where there appears a Y-type flux tube that has a junction,
the flux-tube path length $L$ is defined as the shortest distance along the flux tube 
between two quarks, $Q_i$ and $Q_j$ ($i\neq j$), under consideration.

\subsection{Antitriplet and sextet components}

\begin{figure}[h]
 \includegraphics[width=8cm]{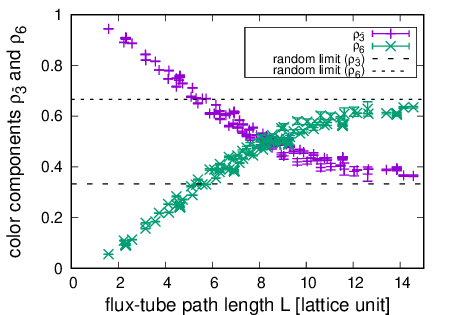}
\caption{\label{Fig.3QComponents}
Antitriplet and sextet components ($\rhot$ and $\rhox$) for 3Q systems plotted as a function of $L$. The dashed and dotted lines indicate the values $\frac39$ and $\frac69$ for $\rhot$ and $\rhox$ in the random-limit color configuration, respectively.}
\end{figure}

Figure~\ref{Fig.3QComponents} shows the $L$ dependence of 
the antitriplet and sextet components $\rhot$ and $\rhox$ that a $QQ$ pair possesses in the 3Q systems.
These components are measured in both of $Q_1Q_2$ and $Q_1Q_3$ pairs in Fig.~\ref{Fig.SPconf3Q},
and they are simultaneously plotted with the same symbols in the figure.
One can find that
a $QQ$ pair in the 3Q system
forms a purely {\it color antitriplet} configuration at $L\rightarrow 0$,
and the ratio $\rhot:\rhox$ 
approaches $3:6$ of the random color configuration as $L$ increases.
Although the color components are measured for various 3Q spatial configurations
(various $\dTQ$ and $\hTQ$),
$\rhot$ and $\rhox$ seem to depend solely on $L$ and are almost single-valued functions of $L$.

From these observations, we can expect that
{\it the $QQ$ color configuration in 3Q systems with ${\cR}\rightarrow 0$ coincides with
the MC color configuration $([Q_1Q_2]Q_{3})$},
and that
{\it the $QQ$ color configuration is randomized when the 3Q system size is enlarged.}
It indicates that the ansatz (Eq.~(\ref{Eq.TAnsatz})) works well for 3Q systems.

As seen in the next subsection, $\rhot$ and $\rhox$
exponentially damp towards the random limit as a function of $L$,
which implies that the quarks' color is screened inside the gluon flux tube connecting two quarks
under investigation.
The exponential damping indicates that in-between flux tube formation is 
expressed by the color leak from color sources to a gluon flux tube,
and can be quantified by the residual rate $\FRt(\cR)$ of the MC color configuration
defined in the ansatz as given in Eq.~(\ref{Eq.TAnsatz}).

\subsection{$L$ dependence of $\FRs(\cR)$}

\begin{figure}[h]
 \includegraphics[width=8cm]{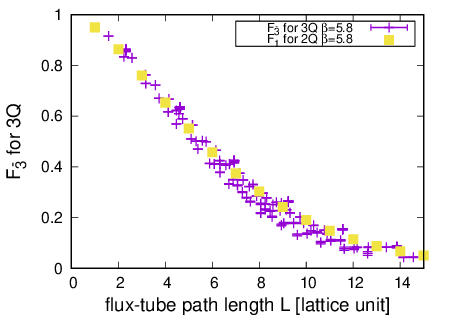}
 \caption{\label{Fig.3QF}
$\QQQFRt(\cR)$ is plotted as a function of $L$.
The square symbols denote $\QQFRs(\cR)$ obtained in a $Q\bar Q$ system.
}
\end{figure}

In Fig.~\ref{Fig.3QF}, the residual rate $\QQQFRt(\cR)$ obtained in 3Q systems is plotted as a function of $L$.
The residual rate $\QQFRs(\cR)$ obtained in $Q\bar Q$ systems denoted by square symbols is also plotted as a function of $L=R$.
$\QQFRs(\cR)$ used here are the same data as shown in Ref.~\cite{Takahashi:2019ghj}.
The positions $\cR$ of the quark and antiquark were set to $\cR = \{\vec r_1,\vec r_2\}=\{(0,0,0),(R,0,0)\}$ in the calculation.
As a remarkable fact, one can find a good coincidence between these two functions $\QQQFRt(\cR)$ and $\QQFRs(\cR)$.
It implies that the color initially associated with quarks is screened ``inside'' the flux tube, and
the screening strength enlarges as the flux tube grows showing the same $L$ dependence
in both of $Q\bar Q$ and 3Q systems.

\section{Lattice QCD results (4Q)}
\label{Sec.Results4Q}

\subsection{spatial configuration}

\begin{figure}[h]
 \includegraphics[width=6cm]{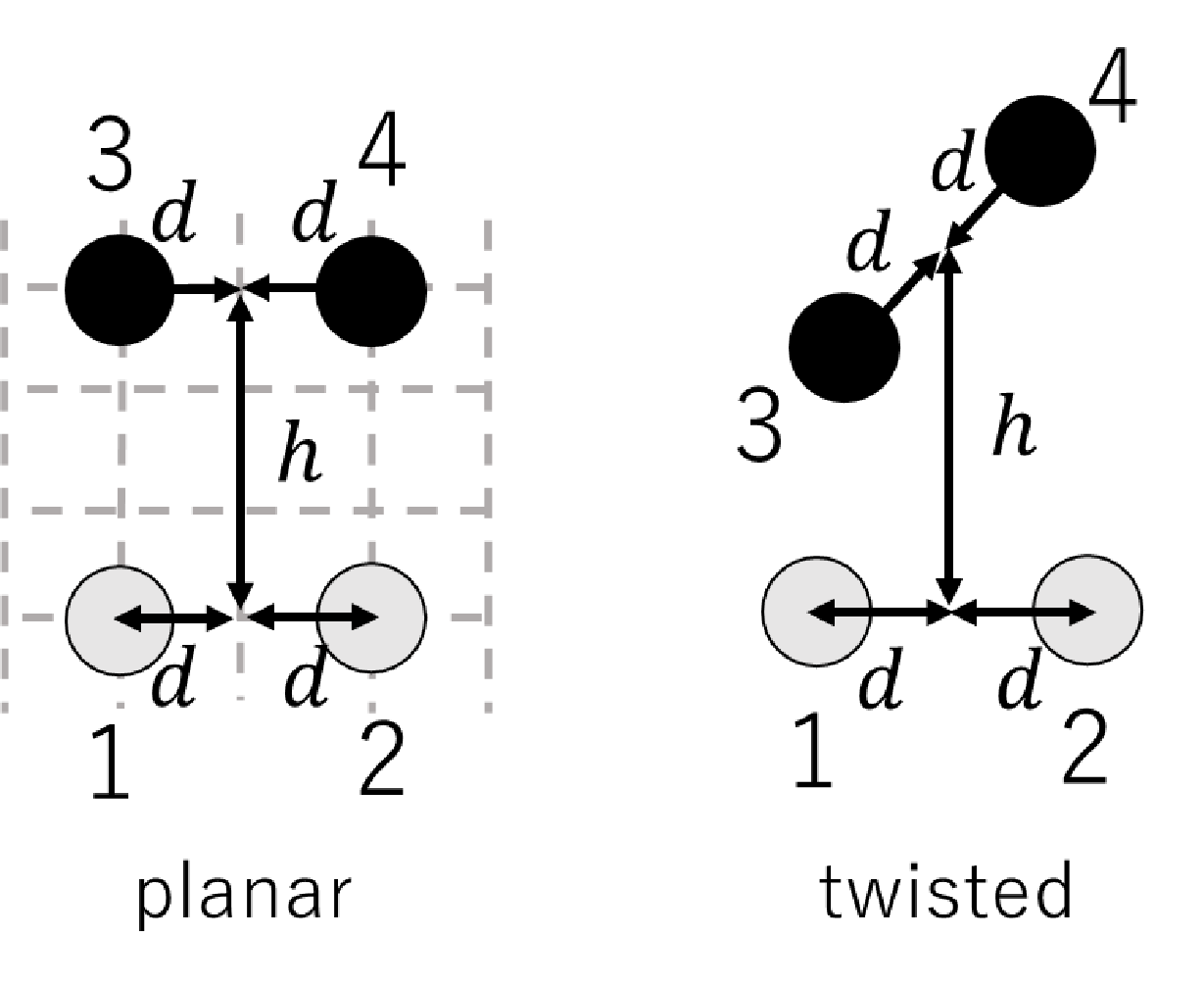}
\caption{\label{Fig.SPconf4Q}
Quarks and antiquarks are located at $(+\dFQ,0,0)$, $(-\dFQ,0,0)$, $(-\dFQ,\hFQ,0)$ and $(+\dFQ,\hFQ,0)$ in the case of planar spatial configurations.
Quarks and antiquarks are located at $(+\dFQ,0,0)$, $(-\dFQ,0,0)$, $(0,\hFQ,-\dFQ)$ and $(0,\hFQ,+\dFQ)$ in the case of twisted spatial configurations.
}
\end{figure}

We put quarks and antiquarks at $(+\dFQ,0,0)$, $(-\dFQ,0,0)$, $(-\dFQ,\hFQ,0)$ and $(+\dFQ,\hFQ,0)$ 
with integers $\dFQ$ and $\hFQ$
in the case of planar spatial configurations,
and quarks and antiquarks are located at $(+\dFQ,0,0)$, $(-\dFQ,0,0)$, $(0,\hFQ,-\dFQ)$ and $(0,\hFQ,+\dFQ)$ in the case of twisted spatial configurations,
as seen in Fig.~\ref{Fig.SPconf4Q}.
We adopt $1\leqq \dFQ \leqq 4$ and $1\leqq \hFQ \leqq 8$ for sets of ($\dFQ$,$\hFQ$),
totally 32 different spatial configurations.

\subsection{Energy spectrum}

\begin{figure}[h]
 \includegraphics[width=7cm]{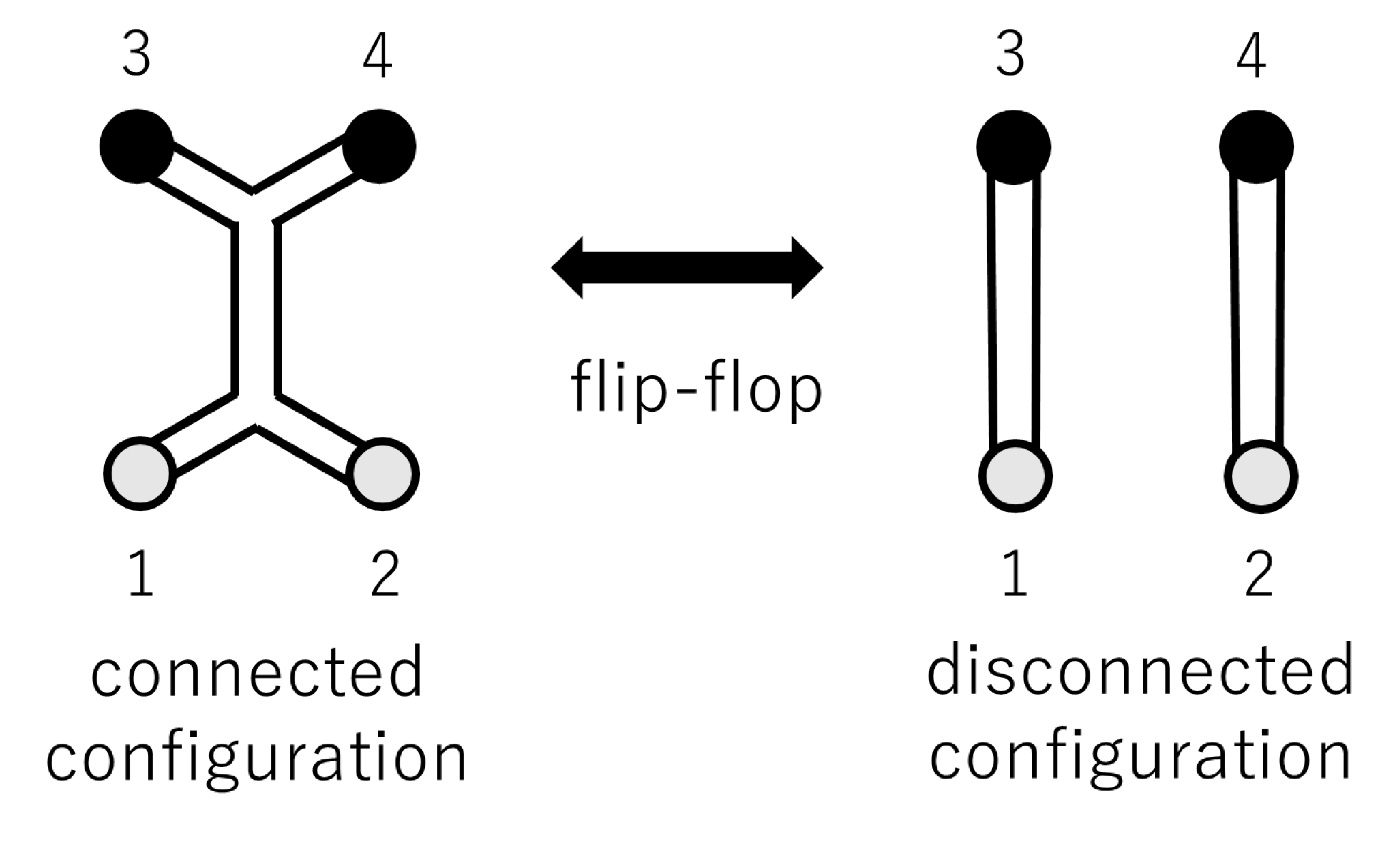}
\caption{\label{Fig.Fform03}
In connected 4Q configurations (left),
all the quarks are connected with a double-Y type flux tube~\cite{Cardoso:2011fq}.
In disconnected 4Q configurations (right),
quark and antiquark are connected with a straight flux tube,
and the system consists of two ``mesons'' (two $Q\bar Q$'s).
These two configurations transform into one another through
the rearrangement of flux tubes called a flip-flop.
}
\end{figure}

The most prominent difference in color structure between twisted and planar systems can be seen
when large $d$ is adopted.
In the case of a planar spatial configuration,
flux-tube rearrangement (flip-flop) occurs as shown in Fig.~\ref{Fig.Fform03},
and both of ``connected 4Q configuration'' and ``disconnected 4Q configuration'' appear
depending on the parameters, $h$ and $d$.
On the other hand, in the case of a twisted spatial 4Q configuration,
the ``connected 4Q configuration'' shown in the left panel in Fig.~\ref{Fig.Fform03}
is always energetically favored~\cite{Okiharu:2004ve},
and no flip-flop occurs.

\begin{figure}[h]
\includegraphics[width=08cm]{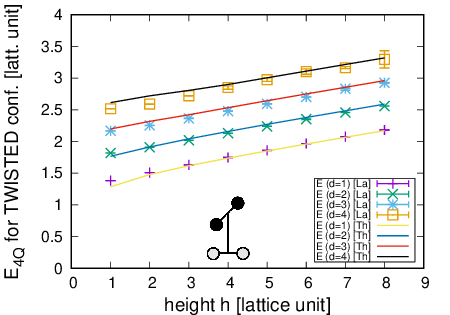}
\caption{\label{Fig.4QSpectraT}
The data with errorbars labeled ``E $(d=i)$ [La]'' show the energy spectra of twisted 4Q configurations for each $\dFQ$
plotted as a function of $\hFQ$.
Solid lines labeled ``E $(d=i)$ [Th]'' show theoretical expectation values
for ``connected'' 4Q configurations
obtained assuming a Coulomb plus double-Y type (X-type) linear confinement potential.}
\end{figure}
\begin{figure}[h]
\includegraphics[width=08cm]{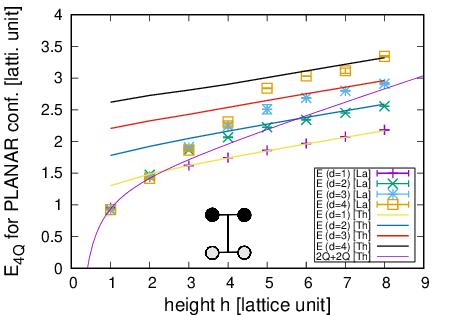}
\caption{\label{Fig.4QSpectraP}
The data with errorbars labeled ``E $(d=i)$ [La]'' show the energy spectra of planar 4Q configurations for each $\dFQ$
plotted as a function of $\hFQ$.
Solid lines labeled ``E ($d=i$) [Th]'' show theoretical expectation values
for ``connected'' 4Q configurations
obtained assuming a Coulomb plus double-Y type (X-type) linear confinement potential.
Solid lines labeled ``2Q+2Q'' show theoretical expectation values
for ``disconnected'' 4Q configurations
assuming that the system is split into ``two mesons'' as a result of flip-flop.}
\end{figure}

In Fig.~\ref{Fig.4QSpectraT},
the data with errorbars labeled ``E $(d=i)$ [La]'' show the energy spectra of twisted 4Q systems for each $d$ plotted as a function of $h$.
Solid lines labeled ``E ($d=i$) [Th]'' show theoretical expectation values
for connected 4Q configurations
that are obtained assuming a Coulomb plus double-Y type linear confinement potential.
Note that the flux-tube profile takes the X-type configuration with large $d$ and small $h$~\cite{Cardoso:2011fq,Okiharu:2004ve},
which is also included in the computation of the theoretical values.
One can find that all the lattice data lie on ``E ($d=i$) [Th]'' curves
and confirm that no flip-flop occurs in twisted 4Q systems.

In Fig.~\ref{Fig.4QSpectraP},
the data with errorbars labeled ``E $(d=i)$ [La]'' represent the energy spectra of planar 4Q systems for each $\dFQ$ plotted as a function of $\hFQ$.
Solid lines labeled ``E ($d=i$) [Th]'' show theoretical expectation values
for connected 4Q configurations
assuming a Coulomb plus double-Y type (X-type) linear confinement potential.
The solid line labeled ``2Q+2Q[Th]'' shows the theoretical expectation value
for disconnected 4Q configurations
obtained assuming that the system is split into ``two mesons'' as a result of flip-flop.
In this case, 4Q energy data approach or lie on a ``2Q+2Q[Th]'' curve at small $\hFQ$.
These observations are consistent with the previous work~\cite{Okiharu:2004ve}.

\subsection{Functional form of the ansatz}

\begin{figure}[h]
 \includegraphics[width=7cm]{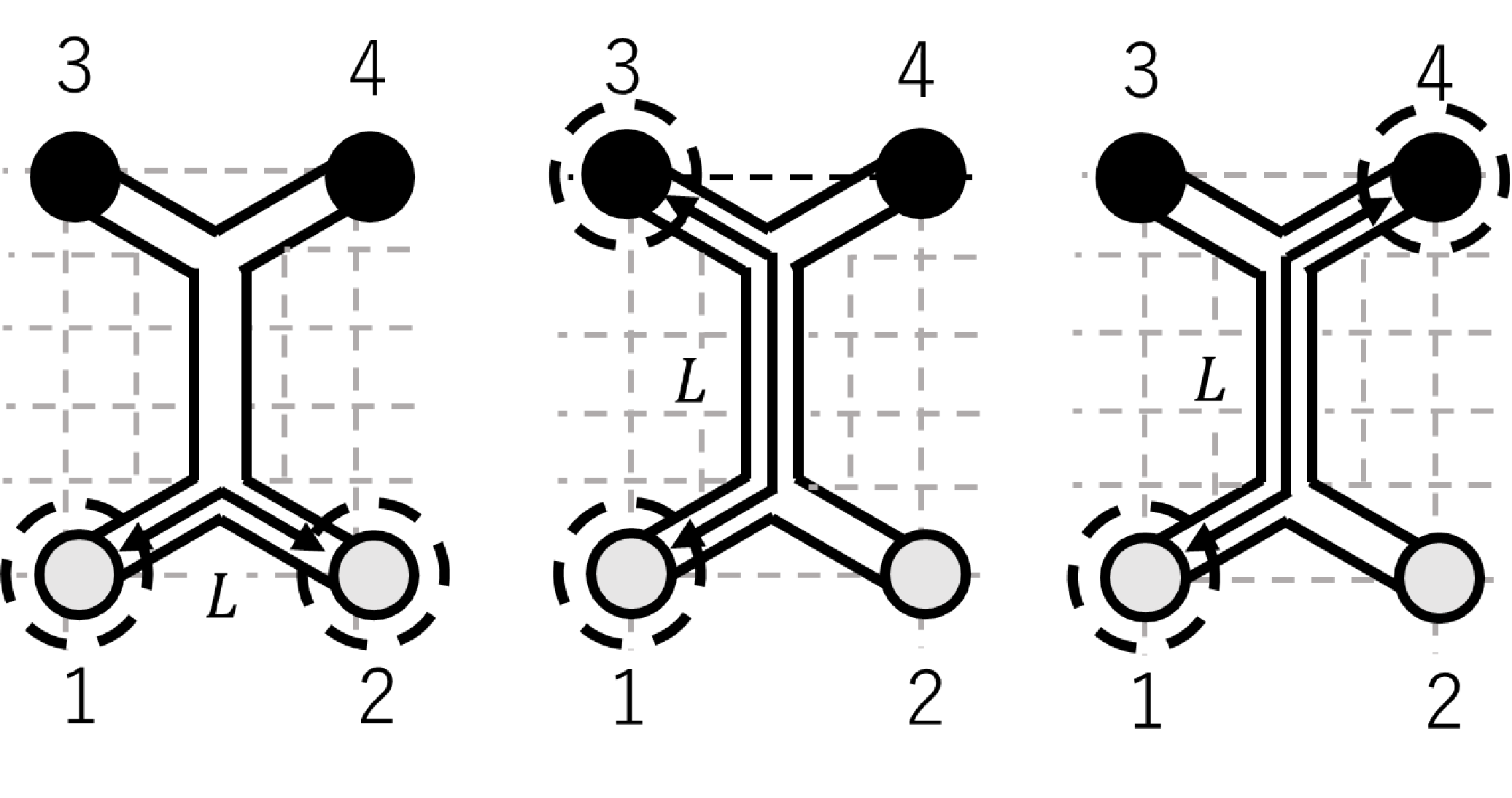}
\caption{\label{Fig.Fform02}
Schematic pictures which express the definition of $L$ for 4Q systems.
Two quarks (antiquarks) of which we measure the color correlation are enclosed in dotted circles.
}
\end{figure}

In the 4Q case, we also evaluate physical quantities based on the flux-tube path length $L$ between two quarks
under consideration.
As for the flux-tube shape, 
we assume a double-Y type (X-type) confining flux tube appearing in ground-state tetraquark systems~\cite{Cardoso:2011fq,Okiharu:2004ve}.
Figure~\ref{Fig.Fform02} shows schematic pictures of $L$ for two quarks (antiquarks) 
enclosed in dotted circles.

Note that planar 4Q systems forming a disconnected configuration ($Q\bar Q$+$Q\bar Q$ state) 
do not have such double-Y type flux tubes,
but they have two separated flux tubes (the right panel in Fig.~\ref{Fig.Fform03}),
where
$L$ can not be defined as a function of $\dFQ$ and $\hFQ$,
hence the color correlation of quark pairs would not exhibit proper $L$
dependence because of flip-flop.

\subsection{$\rhot$ and $\rhos$ in twisted 4Q configurations}

\begin{figure}[h]
 \includegraphics[width=8cm]{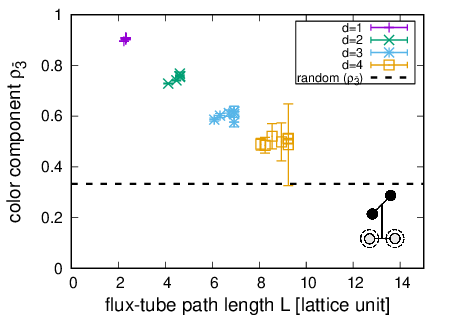}
\caption{\label{Fig.TC_4QT}
Antitriplet components ($\rhot$) for twisted 4Q systems plotted as a function of $L$. The dashed line indicates the random-limit value, $\rhot=3/9$.
}
\end{figure}
\begin{figure}[h]
 \includegraphics[width=8cm]{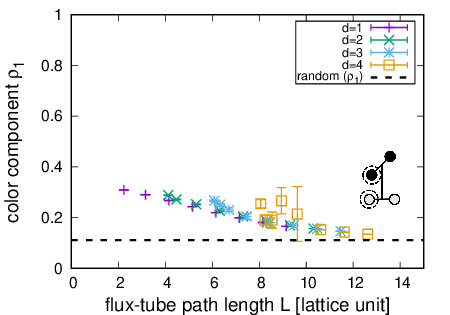}
\caption{\label{Fig.SC_4QT}
Singlet components ($\rhos$) for twisted 4Q systems plotted as a function of $L$. The dashed line indicates the random-limit value, $\rhos=1/9$.
}
\end{figure}

Figure~\ref{Fig.TC_4QT} shows the $L$ dependence of the antitriplet components ($\rhot$) 
in the $Q_1Q_2$ pair (the left panel of Fig.~\ref{Fig.Fform02})
in twisted 4Q spatial configurations,
and Fig.~\ref{Fig.SC_4QT} shows the singlet components ($\rhos$) 
in the $Q_1\bar Q_3$ pair (the middle panel of Fig.~\ref{Fig.Fform02}).
The dashed line in both figures indicates 3/9 and 1/9, which are the random-limit values of $\rhot$ and $\rhos$, respectively.
Note that the density matrix for the $Q_1\bar Q_4$ pair (the right panel of Fig.~\ref{Fig.Fform02}) coincides with that for $Q_1\bar Q_3$
due to the rotational and reflection symmetry on the lattice.

In Figs.~\ref{Fig.TC_4QT} and \ref{Fig.SC_4QT},
we can see that $\rhot$ and $\rhos$ can be expressed by a single function of $L$.
In the case of $\rhot$ for the $QQ$ pair, it starts from $\rhot = 1$ at $L\sim 0$ and gradually approaches $\rhot = 3/9$.
$\rhot = 1$ implies that $QQ$ pair in the ground-state 4Q system forms a purely color triplet state,
and the $QQ$'s color configuration can be expressed purely by $\hrhot$ at $L\sim 0$.
$\rhot$ approaching the random limit value $3/9$ shows that a $QQ$ pair's color configuration is randomized at large $L$ region.
On the other hand, $\rhos$ for the $Q\bar Q$ pair starts from $\rhos = 1/3$ at $L\sim 0$ and gradually approaches
the random-limit value, $\rhos = 1/9$, at large $L$.
$\rhos = 1/3$ implies that $Q\bar Q$ pair's color configuration
in the ground-state 4Q system is a combination of a color singlet and color octet configurations
with the ratio $\rhos:\rhoa=1:2$,
which means that the color state at $L\sim 0$ can be expressed by 
$
\hat\rho^{\mix} = \frac13 \hrhos + \frac23 \hrhoa
$
defined in Eq.~(\ref{Eq.rhomix}) for the $Q\bar Q$ pair in the MC color configuration.
$\rhos$ approaching $1/9$ shows that $Q\bar Q$ pair's color configuration is again randomized at large $L$ region.

From these observations, we can expect that
{\it the quark color configuration for small 4Q systems (${\cR}\rightarrow 0$) coincides with
the MC color configuration $([Q_1Q_2][\bar Q_3 \bar Q_4])$},
and that
{\it the quark color configuration is randomized when the 4Q system size is enlarged as expected.}
Namely, the ansatz (Eqs.~(\ref{Eq.TAnsatz}) and (\ref{Eq.MAnsatz})) works well also for twisted 4Q systems.

\begin{figure}[h]
 \includegraphics[width=8cm]{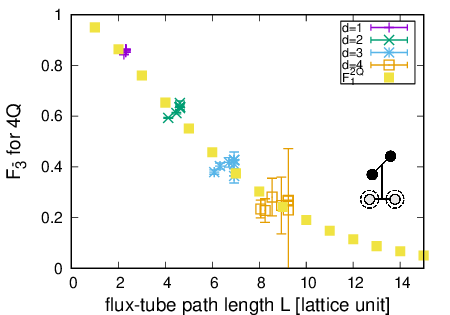}
\caption{\label{Fig.F3_4QT}
The residual rate $\QQQQFRt(\cR)$ in Eq.~(\ref{Eq.TAnsatz}) for twisted 4Q systems plotted as a function of $L$. 
The data labeled ``$F_1^{\rm 2Q}$'' here represents $\QQFRs(\cR)$ obtained in 2Q ($Q\bar Q$) systems.
}
\end{figure}
\begin{figure}[h]
 \includegraphics[width=8cm]{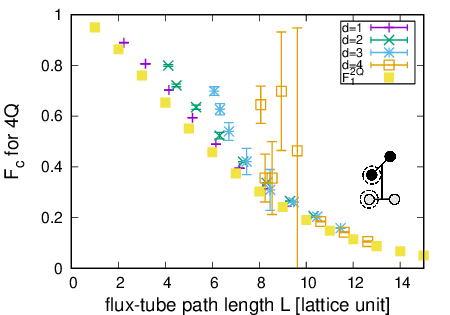}
\caption{\label{Fig.Fm_4QT}
The residual rate $\FRm(\cR)$ in Eq.~(\ref{Eq.MAnsatz}) for twisted 4Q systems plotted as a function of $L$. 
The data labeled ``$F_1^{\rm 2Q}$'' here represents $\QQFRs(\cR)$ obtained in 2Q ($Q\bar Q$) systems.
}
\end{figure}

In Figs.~\ref{Fig.F3_4QT} and ~\ref{Fig.Fm_4QT},
the residual rates of $\rhot$ and $\hat \rho^{\mix}$, $\QQQQFRt(\cR)$ and $\QQQQFRm(\cR)$, are plotted as a function of $L$.
In both figures, the square symbols labeled ``$F_1^{\rm 2Q}$'' represent 
$\QQFRs(\cR)$ obtained in 2Q ($Q\bar Q$) systems.
Figure~\ref{Fig.F3_4QT} shows good agreement in the $L$ dependence between $\QQQQFRt(\cR)$ and $\QQFRs(\cR)$,
which indicates that the color screening occurs inside a gluon flux tube
connecting two quarks ($QQ$) also for 4Q cases,
and its screening effect enlarges along the flux-tube path.
While $\QQQQFRm(\cR)$ globally coincides with $\QQFRs(\cR)$ as seen in Fig.~\ref{Fig.Fm_4QT}, 
one can find some deviation from $\QQFRs(\cR)$ line especially for larger $d$ and smaller $h$.
One possible reason for the deviation might be the flux-tube profile.
When large $d$ and small $h$ are employed, the flux-tube profile approaches the X-type configuration~\cite{Cardoso:2011fq,Okiharu:2004ve}.
In this X-type flux configuration, the number of junctions is one,
and angles between tubes are no longer $\frac23 \pi$.
It is in contrast to the double-Y type configuration,
in which there appear two junctions
on the flux-tube path between $Q$ and $\bar Q$,
and angles between tubes all remain $\frac23 \pi$.

\subsection{$\rhot$, $\rhosa$ and $\rhosb$ in planar 4Q configurations}

\begin{figure}[h]
 \includegraphics[width=8cm]{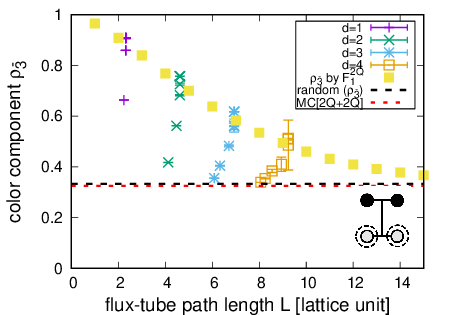}
\caption{\label{Fig.TC_4QP}
Antitriplet components ($\rhot$) for planar 4Q systems plotted as a function of $L$. 
The black dashed line indicates 3/9 representing the random limit value of $\rhot$.
The red dotted line labeled ``MC[2Q+2Q]'' shows 3/9, which is the value of $\rhot$ expected when a system forms a ``disconnected 4Q configuration''.
The data labeled ``$\rhot$ by $F_1^{\rm 2Q}$'' here represents the reconstructed values of $\rhot$ 
when $\QQFRs(\cR)$ for $Q\bar Q$ systems is inserted into $\FRt(\cR)$ in Eq.~(\ref{Eq.TAnsatz}).}
\end{figure}
\begin{figure}[h]
 \includegraphics[width=8cm]{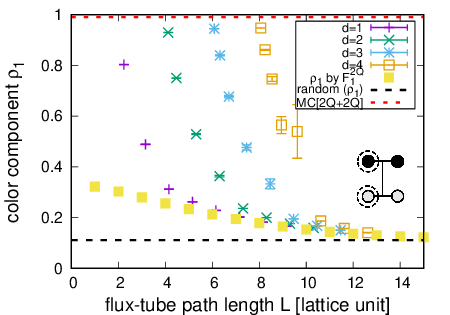}
\caption{\label{Fig.SCa_4QP}
Singlet components ($\rhosa$) for planar 4Q systems plotted as a function of $L$. 
The black dashed line indicates 1/9 representing the random limit value of $\rhos$.
The red dotted line labeled ``MC[2Q+2Q]'' shows 1, which is the value of $\rhosa$ expected when a system forms a ``disconnected 4Q configuration'' and each $Q\bar Q$ pair forms the MC configuration (pure color singlet configuration).
The data labeled ``$\rhos$ by $F_1^{\rm 2Q}$'' here represents the reconstructed values of $\rhos$ 
when $\QQFRs(\cR)$ for $Q\bar Q$ systems is inserted into $\FRm(\cR)$ in Eq.~(\ref{Eq.MAnsatz}).}
\end{figure}
\begin{figure}[h]
 \includegraphics[width=8cm]{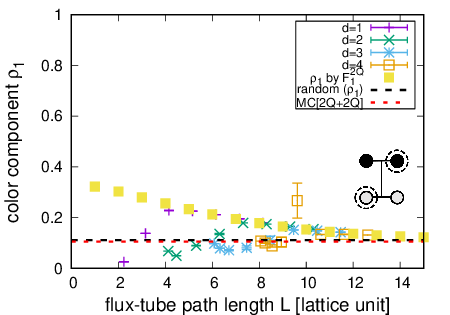}
\caption{\label{Fig.SCb_4QP}
Singlet components ($\rhosb$) for planar 4Q systems plotted as a function of $L$. 
The black dashed line indicates 1/9 representing the random limit value of $\rhos$.
The red dotted line labeled ``MC[2Q+2Q]'' shows 1/9, which is the value of $\rhosb$ expected when a system forms a ``disconnected 4Q configuration''.
The data labeled ``$\rhos$ by $F_1^{\rm 2Q}$'' here represents the reconstructed values of $\rhos$ 
when $\QQFRs(\cR)$ for $Q\bar Q$ systems is inserted into $\FRm(\cR)$ in Eq.~(\ref{Eq.MAnsatz}).}
\end{figure}

Figure~\ref{Fig.TC_4QP} shows 
the antitriplet components ($\rhot$) for a $Q_1Q_2$ pair in planar 4Q systems 
plotted as a function of the flux-tube path length $L$. 
The black dashed line indicates the random limit value 3/9 of $\rhot$.
The red dotted line labeled ``MC[2Q+2Q]'' shows 
$\rhot = 3/9$, the value expected in the MC limit of a ``disconnected 4Q configuration''.
This value coincides with the random-limit value ($\rhot=3/9$),
since in disconnected 4Q configurations,
$Q_1$ and $Q_2$ belong to different color singlet $Q\bar Q$ clusters,
and have no color correlation.
The data labeled ``$\rhot$ by $F_1^{\rm 2Q}$'' here represent the reconstructed values of $\rhot$ 
when $\QQFRs(\cR)$ for $Q\bar Q$ systems is inserted into $\FRt(\cR)$ in Eq.~(\ref{Eq.TAnsatz}).

Figure~\ref{Fig.SCa_4QP} shows 
the singlet components ($\rhosa$) measured in a $Q_1\bar Q_3$ pair in planar 4Q systems 
plotted as a function of the flux-tube path length $L$. 
The black dashed line indicates the random limit value, $\rhosa=1/9$.
The red dotted line labeled ``MC[2Q+2Q]'' shows
$\rhos=1$, the value expected in the MC limit of a disconnected 4Q configuration.
In disconnected 4Q configurations, $Q_1$ and $\bar Q_3$ belong to the identical $Q\bar Q$ cluster,
and $Q_1\bar Q_3$ color correlation would be maximized leading to $\rhos\sim 1$
when $Q_1\bar Q_3$ distance $h$ is small.
The data labeled ``$\rhos$ by $F_1^{\rm 2Q}$'' here represents the reconstructed values of $\rhos$ 
when $\QQFRs(\cR)$ for $Q\bar Q$ systems is inserted into $\FRm(\cR)$ in Eq.~(\ref{Eq.MAnsatz}).

Figure~\ref{Fig.SCb_4QP} shows 
the singlet components ($\rhosb$)
measured in a $Q_1\bar Q_4$ pair in planar 4Q systems 
plotted as a function of the flux-tube path length $L$. 
The black dashed line indicates the random limit value, $\rhosb=1/9$.
The red dotted line labeled ``MC[2Q+2Q]'' shows $\rhos=1/9$, the value expected in the MC limit of a disconnected 4Q configuration.
In disconnected 4Q configurations, $Q_1$ and $\bar Q_4$ belong to different color singlet $Q\bar Q$ clusters,
and have no color correlation leading to $\rhos\sim 1/9$.
The data labeled ``$\rhos$ by $F_1^{\rm 2Q}$'' represents the reconstructed values of $\rhos$ 
when $\QQFRs(\cR)$ for $Q\bar Q$ systems is inserted into $\FRm(\cR)$ in Eq.~(\ref{Eq.MAnsatz}).

In planar 4Q systems, systems would form a connected 4Q configuration at large $h$.
With a decrease of $h$, the energy of a ``two-meson state'' comes down lower than that of genuine connected 4Q states,
and the system will undergo flip-flop and form a disconnected 4Q state at small $h$.
In such disconnected 4Q cases, if we measure $\rhot$ or $\rhosb$ in 4Q systems, 
it would take the random-limit value $\rhot=3/9$ ($\rhosb=1/9$),
since $Q_1$ and $Q_2$($\bar Q_4$) under investigation belong to different color singlet clusters and have no color correlation.
Even in connected 4Q systems, $\rhot$ and $\rhosb$ also approach the random limit values
($\rhot=3/9$, $\rhosb=1/9$) in the large $L$ limit
due to the color screening inside a flux tube,
but the $L$ dependence of $\rhot$ and $\rhosb$ in this case would be on the expected curve as shown later,
and we can discriminate these behaviors.
When we measure $Q_1\bar Q_3$ color correlation in disconnected 4Q systems with small $h$,
$\rhosa$ would be 1,
since $Q_1$ and $\bar Q_3$ under investigation belong to the same color singlet cluster
and have maximal color correlation.

The lattice data $\rhot$ in Fig.~\ref{Fig.TC_4QP} 
for small $\hFQ$ lie around $\rhot=3/9$ (random limit value), which indicates that 
there is no color correlation between two quarks ($Q_1Q_2$) under investigation and
systems are just forming disconnected 4Q configurations.
As $\hFQ$ increases, the data approach the values 
reconstructed from $\QQFRs(\cR)$ for $Q\bar Q$ systems (filled square symbols in the figures),
which shows that the systems are genuine connected 4Q states for large $\hFQ$,
and $\rhot$ shows the proper $L$ dependence that is expected for the connected 4Q case.

The lattice data $\rhosa$ in Fig.~\ref{Fig.SCa_4QP} 
for small $\hFQ$ lie around $\rhosa=1$, which indicates that 
$Q_1\bar Q_3$ under investigation are forming a genuine color-singlet pair
and shows that systems are divided into two-meson clusters
in a disconnected 4Q state.
As $\hFQ$ increases, the data approach the values 
reconstructed from $\QQFRs(\cR)$ for $Q\bar Q$ systems (filled square symbols in the figures),
which again confirms that the systems are genuine connected 4Q states for large $h$.

The lattice data $\rhosb$ in Fig.~\ref{Fig.SCb_4QP} 
for small $h$ lie around $\rhosb=1/9$, which indicates that 
$Q_1\bar Q_4$ under investigation have no color correlation and belong to different color-singlet clusters,
and that systems are 4Q disconnected states.
As $\hFQ$ increases, the data approach the values 
reconstructed from $\QQFRs(\cR)$ for $Q\bar Q$ systems (filled square symbols in the figures),
which repeatedly shows that the systems are genuine connected 4Q states for large $\hFQ$.
In these analyses, the data for $d=1$ and small $h$ show slightly unexpected behaviors,
which is considered to reflect possible remaining interactions between two-mesons.

Based on the analyses of 2Q, 3Q, and 4Q systems, we find the following.
\begin{itemize}
 \item 
When a system size is small and no gluon flux tube is present, 
quarks form the ``maximally correlated (MC)'' color configuration,
which is naively expected when we ignore gluon's color.
 \item 
As a system size is enlarged and a long gluon flux tube grows,
quarks' color is screened inside the flux tube,
and finally the quarks' color configuration approaches the ``random'' color configuration,
in which all the color components equally contribute,
as a result of color screening.
 \item 
The color configuration of any quark pair can be represented by the color density operator
\[
\hat\rho^{\rm ansatz}({\cR})
=
F_{\rm MC}({\cR})\hat\rho^{\rm MC} + (1-F_{\rm MC}({\cR}))\hat\rho^{\rm rand}.
\]
$\hat\rho^{\rm MC}$ and $\hat\rho^{\rm rand}$ correspond to
the operators for the MC color configuration and the random configuration, respectively.
$F_{\rm MC}({\cR})$ is the residual rate of the MC color configuration,
which is a monotonous function of a flux-tube path length $L$.
 \item 
The flux-tube path length $L$ dependence of the residual rates 
of the MC color configurations
(${\QQFRs({\cR})}$, ${\QQQFRt({\cR})}$, ${\QQQQFRt({\cR})}$ and ${\QQQQFRm({\cR})}$) 
all show {\it universality};
the universal $L$ dependence of the color screening effect along the flux-tube path.
 \item 
When two quarks in a pair under consideration belong to different color singlet clusters (mesons),
they have no color correlation and
their color configuration is expressed by the random color configuration,
which can be useful to identify the internal structure of multiquark states.
\end{itemize}

\section{Analysis with Entanglement Entropy}
\label{Sec.ResultsEE}

Entanglement entropy (EE)~\cite{Takahashi:2019ghj,Takahashi:2020bje,Calabrese:2004eu,Ryu:2006ef,Kitaev:2005dm,Amico:2007ag,Solodukhin:2011gn,Donnelly:2014gva,Amorosso:2024leg}
defined by a reduced density matrix
reflects and quantifies
the correlation between the d.o.f. in the reduced density matrix and the traced-out d.o.f. in a full density matrix.
Let us consider a system which is divided into subsystems $A$ and $B$.
Defining \[
\rho_A\equiv {\rm Tr}_B \ \left(\rho_{\rm full}\right),
    \]
a reduced density matrix obtained by tracing out the d.o.f. for the subsystem $B$,
an entanglement entropy $S^{\rm EE}$ in the functional form of the von Neumann entropy
is expressed as
\begin{equation}
S^{\rm EE} = -{\rm Tr}_A \left(\rho_A\log \rho_A\right)\equiv S^{\rm vN},
\end{equation}
and EE in the form of the Renyi entropy of the order $\alpha$ can be expressed as
\begin{equation}
S^{\rm EE} = \frac{1}{1- \alpha}\log {\rm Tr}_A \left(\rho_A^\alpha \right)\equiv S^{{\rm Renyi}-\alpha}.
\end{equation}
$S^{\rm EE}$ represents the correlation between subsystems $A$ and $B$,
and $S^{\rm EE}\sim 0$ ($S^{\rm EE} \gg 0$) holds when the correlation 
between $A$ and $B$ is weak (strong).

\begin{figure}[h]
 \includegraphics[width=7cm]{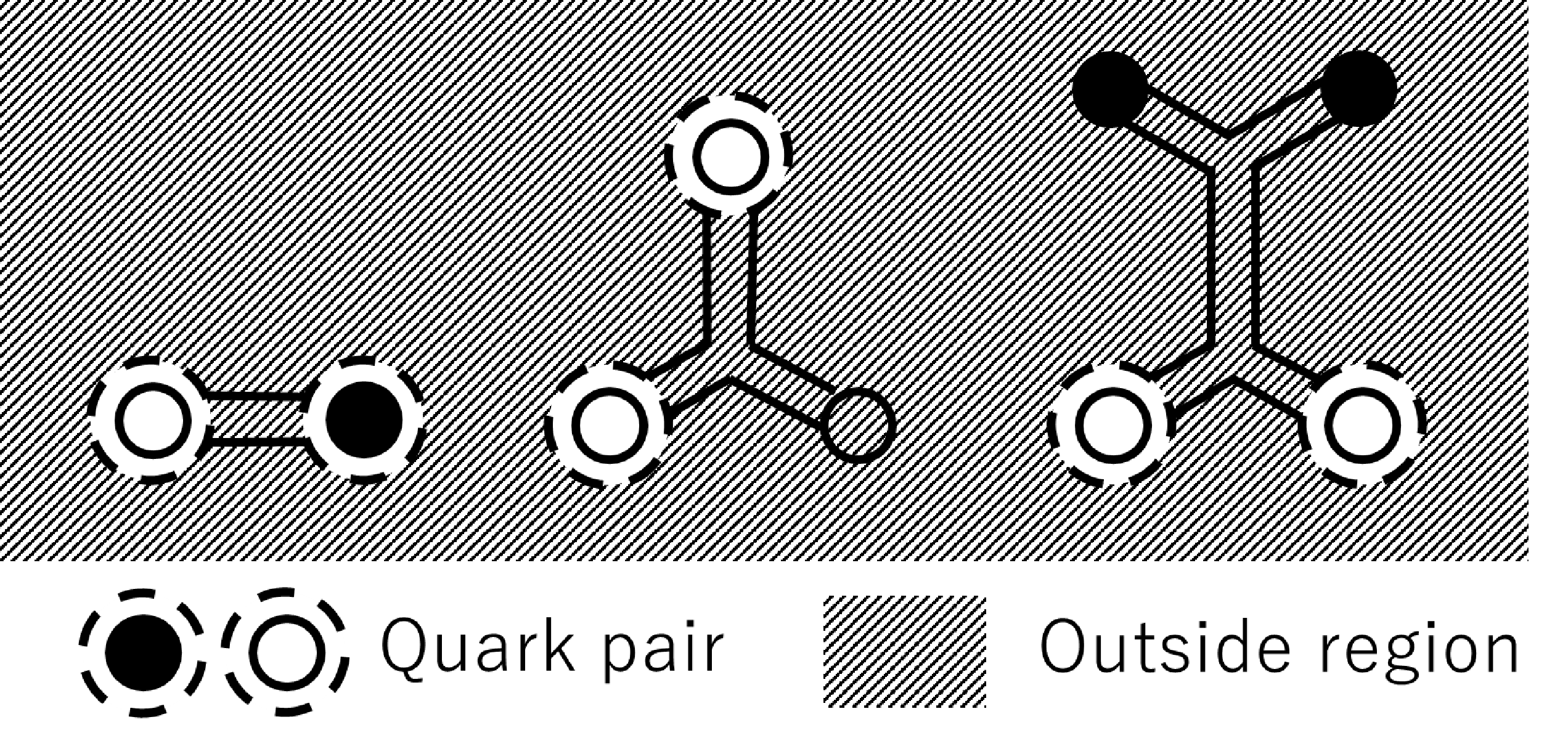}
\caption{\label{Fig.INOUT}
Two quarks in a pair we investigate are enclosed in dotted circles.
The ``outside region'' contains in-between gluon fields and spectator quarks,
whose d.o.f. are all integrated out in the construction of the reduced two-body density matrix $\rho$.
}
\end{figure}

Quark color correlations can be also quantified
from the viewpoint of an entanglement entropy (EE)~\cite{Takahashi:2019ghj,Takahashi:2020bje}.
EE constructed from a reduced two-body color density matrix $\rho$,
which is obtained by integrating out the d.o.f. other than a target quark pair,
can quantify the color correlation between the quark pair and the ``outside region''.
The ``outside region'' here contains in-between gluon fields and all the spectator quarks,
whose d.o.f. are all integrated out in the construction of the reduced two-body density matrix $\rho$.
The situation is illustrated in Fig.~\ref{Fig.INOUT}.
The quark pair of which we investigate the correlation belongs to the subsystem $A$,
and the other d.o.f. including gluons and spectator quarks are all included in the subsystem $B$
in the above example.

In the 2Q ($Q\bar Q$) case,
when no physical gluon flux tube is present in a system,
the quark pair's color does NOT leak into the ``outside region'':
the color flowing out from one (anti)quark is fully absorbed by the other (anti)quark,
and the color d.o.f. of the quark pair decouples from the outside region.
As a result, the color correlation between quarks of the pair is maximized,
and the entanglement (color correlation) between the quark pair and ``outside region'' is zero,
which leads to the minimal value of EE ($S\sim 0$).
When long gluon flux tubes are formed in a system,
quark pair's color largely leak into the flux tube, ``outside region''.
This color screening in the flux tube weakens the interquark color correlation,
and gives a strong entanglement between the quark pair and ``outside region''
providing a large value of EE ($S\gg 0$).
The quark pair's color configuration is randomized
due to the screening by in-between gluons when a flux tube is formed between quarks.

In the case of 3Q ($QQQ$) and 4Q ($QQ\bar Q\bar Q$) cases,
the color from a target quark pair can be absorbed also by the spectator quark(s),
which produces a finite color correlation between the pair and the ``outside region'',
and it gives a finite EE ($S>0$) even when no physical gluon flux tube is present.
EE can quantify the internal color structure in this way, and
we utilize an entanglement entropy to cast light on the internal color structure of multiquark systems.

In this study, we employ the second order Renyi entanglement entropy by taking $\alpha=2$, 
which is simply given by the squared $\rho({\cR})$ as  
\begin{equation}
S^{{\rm Renyi}-2}({\cR}) = -\log {\rm Tr}\left(\rho({\cR})^2 \right).
\end{equation}

\subsection{EE for planar 4Q configurations}

\begin{figure}[h]
 \includegraphics[width=8cm]{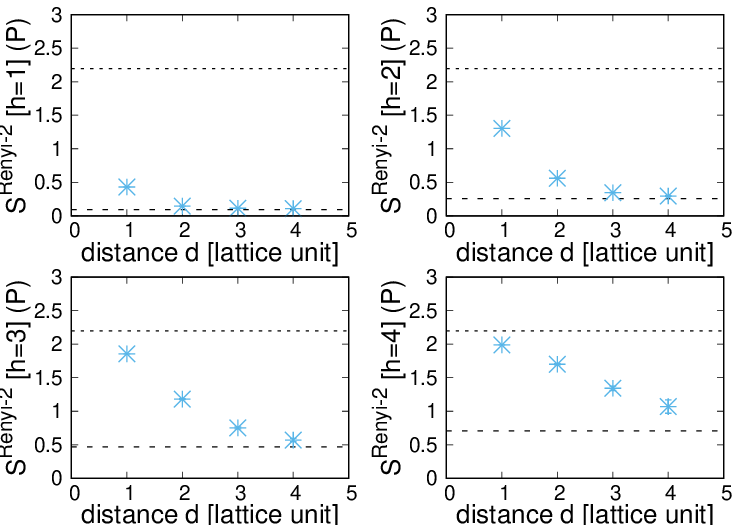}
\caption{\label{Fig.Renyi_14}
$S^{\rm Renyi-2}$ for planar 4Q configurations with $\hFQ=1\sim 4$,
which are measured with $Q_1$ and $\bar Q_3$, are plotted as a function of $\dFQ$.
The upper line denotes $S^{\rm Renyi-2}=\log(N_c^2)$, 
which holds when $Q_1$ and $\bar Q_3$ form a purely random-color state
and have no color correlation.
The lower line denotes $S^{\rm Renyi-2}$ value
when $Q_1$ and $\bar Q_3$ belong to the identical color-singlet cluster
forming a meson state with the interquark distance $R=\hFQ$,
which was separately measured with same parameters in a $Q\bar Q$ system.
}
\end{figure}

\begin{figure}[h]
 \includegraphics[width=8cm]{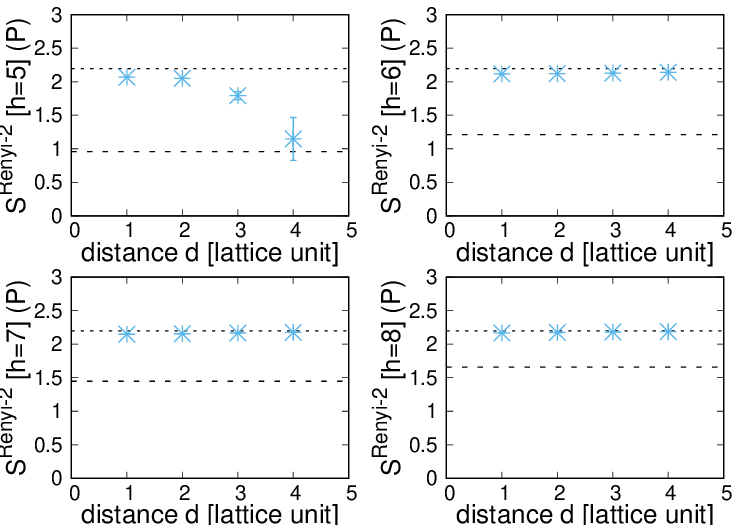}
\caption{\label{Fig.Renyi_58}
$S^{\rm Renyi-2}$ for planar 4Q configurations with $\hFQ=5\sim 8$,
which are measured with $Q_1$ and $\bar Q_3$, are plotted as a function of $\dFQ$.
The upper line denotes $S^{\rm Renyi-2}=\log(N_c^2)$, 
which holds when $Q_1$ and $\bar Q_3$ form a purely random-color state
and have no color correlation.
The lower line denotes $S^{\rm Renyi-2}$ value
when $Q_1$ and $\bar Q_3$ belong to the identical color-singlet cluster
forming a meson state with the interquark distance $R=\hFQ$,
which was separately measured with same parameters in a $Q\bar Q$ system.
}
\end{figure}

In Figs.~\ref{Fig.Renyi_14} and \ref{Fig.Renyi_58},
$S^{\rm Renyi-2}$ for planar 4Q configurations with $\hFQ=1\sim 8$,
which are measured with $Q_1$ and $\bar Q_3$, are plotted as a function of $\dFQ$.
The upper line denotes the maximal value $S^{\rm Renyi-2}=\log(N_c^2)$ 
for a $Q_1\bar Q_3$ pair forming a purely random-color state and having no color correlation.
We again note that {\it no correlation} inside the target pair
implies a {\it maximal correlation} between the pair and ``outside region''.
We also show the $S^{\rm Renyi-2}$ value of an isolated $Q\bar Q$ system (a meson state)
with the interquark distance $R=\hFQ$ by the lower line,
which was separately calculated with the same parameters in a $Q\bar Q$ system.

In three panels for $h=1\sim 3$,
$S$ values are consistent with the two-meson state values denoted by the lower lines at large $d$ region, $d>h$.
It indicates that $Q_1$ and $\bar Q_3$ quarks with a large $d$ and small $h$
in the planar 4Q configuration
belong to the same color singlet $Q_1\bar Q_3$ cluster,
reflecting the fact that the four quarks are separated into two meson clusters
forming a disconnected 4Q configuration
as a result of flip-flop (the right panel in Fig.~\ref{Fig.Fform03}).
In this case, the color from the $Q_1\bar Q_3$ pair is absorbed and screened only by in-between flux tube,
and does not flow out from the $Q_1\bar Q_3$ cluster.

On the other hand, in the panels for $h=6\sim 8$ in Fig.~\ref{Fig.Renyi_58}, 
$S$ lie on the upper line denoting the maximal value of $S$,
which shows that $Q_1\bar Q_3$ pair's color largely leaks into the ``outside region''.
It implies that there exist long flux tube connecting $Q_1$ and $\bar Q_3$,
and that the system would be a genuine connected 4Q state (the left panel in Fig.~\ref{Fig.Fform03})
in the large $h$ region.
In this connected 4Q case, the color of $Q_1\bar Q_3$ pair also flows into spectator quarks $Q_2$ and $\bar Q_4$ as well as in-between flux tube,
which further enlarges $S$.

The 4Q states whose EE values lie between the two reference lines
are considered to be in the transient stage of color rearrangement (flip-flop) process.

\subsection{EE for twisted 4Q configurations}

\begin{figure}[h]
 \includegraphics[width=8cm]{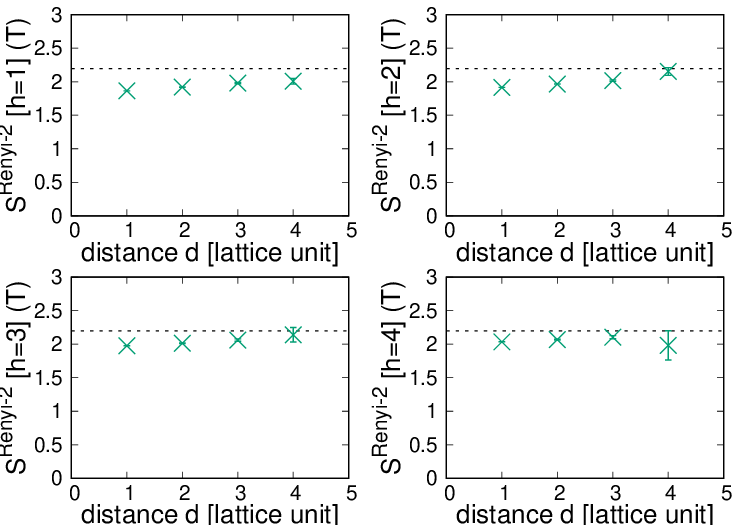}
\caption{\label{Fig.Renyi_14T}
$S^{\rm Renyi-2}$ for twisted 4Q configurations with $\hFQ=1\sim 4$,
which are measured with $Q_1$ and $\bar Q_3$, are plotted as a function of $\dFQ$.
The dotted line denotes $S^{\rm Renyi-2}=\log(N_c^2)$, the maximum value of EE.
}
\end{figure}

\begin{figure}[h]
 \includegraphics[width=8cm]{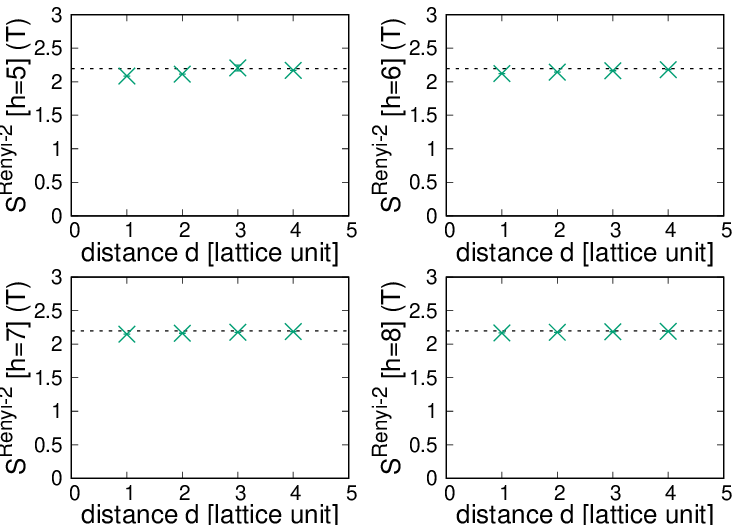}
\caption{\label{Fig.Renyi_58T}
$S^{\rm Renyi-2}$ for twisted 4Q configurations with $\hFQ=5\sim 8$,
which are measured with $Q_1$ and $\bar Q_3$, are plotted as a function of $\dFQ$.
The dotted line denotes $S^{\rm Renyi-2}=\log(N_c^2)$, the maximum value of EE.
}
\end{figure}

In Figs.~\ref{Fig.Renyi_14T} and \ref{Fig.Renyi_58T},
$S^{\rm Renyi-2}$ for twisted 4Q configurations with $\hFQ=1\sim 8$,
which are measured with $Q_1$ and $\bar Q_3$, are plotted as a function of $\dFQ$.
The dotted line denotes $S^{\rm Renyi-2}=\log(N_c^2)$, the maximal value of EE.
In stark contrast to the planar 4Q case,
EE takes large values close to $\log(N_c^2)$ even with small $h$ and $d$,
and the large EE in these regions implies large color leak from a quark pair to the ``outside region''.
Taking into account that in-between flux tube has not grown well in these small-size regions,
the quark pair's color does not leak to in-between gluons but partially flows to other ``spectator'' quarks ($Q_2$ and $\bar Q_4$),
which again verifies that four quarks are  ``connected'' in terms of color in twisted 4Q systems.

\subsection{EE and flip-flop}

\begin{figure}[h]
 \includegraphics[width=8cm]{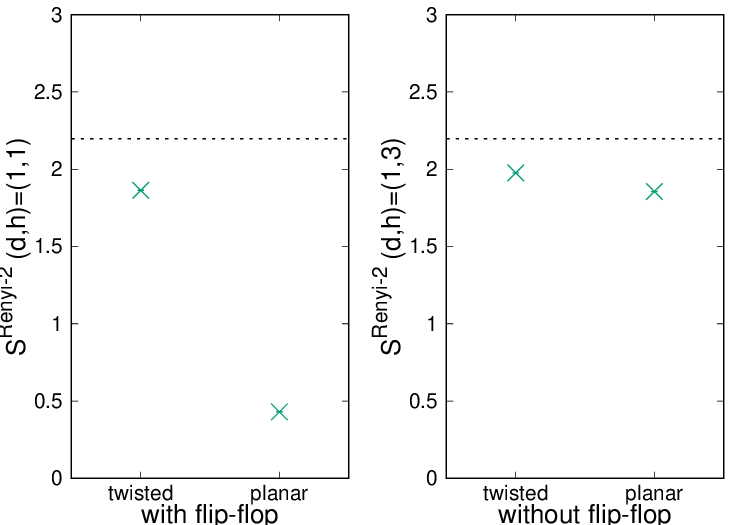}
\caption{\label{Fig.FF}
$S^{\rm Renyi-2}$ for twisted and planar configurations with $(d,h)=(1,1)$ and $(1,3)$,
which are measured with $Q_1$ and $\bar Q_3$, are shown.
The dotted line denotes $S^{\rm Renyi-2}=\log(N_c^2)$, the maximum value of EE.
Flip-flop by twisting occurs with $(d,h)=(1,1)$, and not for $(d,h)=(1,3)$.
}
\end{figure}

In Fig.~\ref{Fig.FF}, 
$S^{\rm Renyi-2}$ for twisted and planar configurations with $(d,h)=(1,1)$ (left panel)
and $(d,h)=(1,3)$ (right panel) are compared.
Flip-flop due to twisting occurs with $(d,h)=(1,1)$, and not for $(d,h)=(1,3)$,
which is confirmed based on the analysis conducted thus far.
As can be seen in both panels, the flip-flop, the rearrangement of color structure, 
can be clearly observed as a large change in EE.
An entanglement entropy $S$ constructed from a two-body reduced color density matrix $\rho$
is expected to cast light on the internal color confinement structure.

\section{Summary and concluding remarks}
\label{Sec.Summary}

  We have investigated the color correlation between two 
  quarks in static multiquark ($N$Q) systems
  in the confined phase by lattice QCD.
  We have performed quenched lattice QCD calculations with the Coulomb gauge
  adopting the standard Wilson gauge action,
  and the spatial volume considered here is $L^3 = 32^3$ at $\beta = 5.8$,
  which corresponds to the lattice spacing $a=0.14$ fm 
  and the system volume $L^3=4.5^3$ fm$^3$.
  The two-body reduced density matrix $\rho$ in the color space
  has been constructed from link variables. The calculated $\rho$ has been analyzed based on the ansatz
  in which the color leak into gluon flux tubes is expressed by mixing of the random density matrix $\rho^{\rm rand}$ as
  we proposed in our previous papers~\cite{Takahashi:2019ghj,Takahashi:2024vff}.

  Through the whole analyses of 2Q, 3Q and 4Q systems,
  we have
  clarified the following issues.

{\it Quarks' color structure for small system-size limit:}
When a system size is small and no gluon flux tube is present, 
quarks form the ``maximally correlated (MC)'' color configuration,
which is naively expected when we ignore gluon's color.

{\it Quarks' color screening by inbetween gluon flux tubes:}
As a system size is enlarged and a long gluon flux tube grows,
quarks' color is screened inside the flux tube,
and finally the quarks' color configuration approaches the ``random'' color configuration,
in which all the color components equally contribute,
as a result of color screening.

{\it Density operator $\hat\rho$ that reproduces quarks' color configurations:}
The color configuration of any quark pair can be represented by the color density operator
\[
\hat\rho^{\rm ansatz}({\cR})
=
F_{\rm MC}({\cR})\hat\rho^{\rm MC} + (1-F_{\rm MC}({\cR}))\hat\rho^{\rm rand}.
\]
$\hat\rho^{\rm MC}$ and $\hat\rho^{\rm rand}$ correspond to
the operators for the MC color configuration and the random configuration, respectively.
$F_{\rm MC}({\cR})$ is the residual rate of the MC color configuration,
which is a monotonous function of a flux-tube path length $L$.

{\it Universality of residual rates $F$:}
The flux-tube path length $L$ dependence of the residual rates 
of the MC color configurations
(${\QQFRs({\cR})}$, ${\QQQFRt({\cR})}$, ${\QQQQFRt({\cR})}$ and ${\QQQQFRm({\cR})}$) 
show {\it universality};
the universal $L$ dependence of the color screening effect along the flux-tube path.

{\it Quarks' color structure when a system is divided into multi color-singlet clusters:}
When two quarks in a pair under consideration belong to different color singlet clusters (mesons),
they have no color correlation and
their color configuration is expressed by the random color configuration,
which can be useful to identify the internal structure of multiquark states.

{\it Property of entanglement entropy $S$:}
The entanglement entropy $S$ defined from the two-body color density matrix $\rho$ can probe
the rearrangement of the flux tubes and internal color structure in multiquark systems,
such as the flip-flop process in 4Q systems.

We comment on the gauge dependence of our findings.
In the present calculation,
we employ the Coulomb gauge because the gauge fixing is needed
to measure the quarks' color correlations.
Since the present analysis is based on 
gauge dependent variables, 
the obtained results would be more or less gauge dependent.
Nevertheless, 
we believe that our present results can cast light on the confinement physics
in multiquarks systems.
Indeed, our method in the Coulomb gauge 
can probe the flip-flop signal inside multiquark systems,
which is surely a gauge invariant ``physical'' process.
Moreover, in the present results,
the color correlations between quarks measured in the Coulomb gauge 
actually depends on the physical multi-junction flux tube profile (flux tube path length $L$), which is also extracted with no gauge dependence.
In these senses, we expect that our methodology does not only give a picture in the Coulomb gauge,
but gives a physical implication for multiquark systems.
The comparison of the present results with those
in other possible gauges is a remaining problem.

The present results of the $L$ (flux tube path length) dependence of quarks' color correlations
and the universality of the damping rate (universality of the color screening mass)
indicate that the quark pair color screening in flux tubes
occurs in a unified manner even in multiquark systems,
in which the target quark pair is surrounded by other spectator quarks and gluons,
and support the multi-junction flux tube formation 
as well as the robustness of the flux tubes inside multiquark systems.
Moreover, our prescription can clearly probe the flip-flop signal, 
the rearrangement of flux tubes in multiquark systems, 
which is one of the great advantages of our method.
Possible future directions of this research would be application of our method
to the study of interquark color correlations
as well as color structures and screening effects of flux tubes
inside hadrons with dynamical (not static) quarks, inclusion of dynamical sea quarks, etc.
It is also a future problem to utilize our method for researches of the internal structures
of excited hadrons or multiquark exotic hadrons in a nonperturbative manner.

\begin{acknowledgements}
This work was partly supported by
Grants-in-Aid of the Japan Society for the Promotion of
Science (Grant Nos. 18H05407, 22K03608, 22K03633, 25K07310),
and was partly achieved through the use of SQUID at D3 center, Osaka University.
The computational resources were provided by RCNP, Osaka University.
\end{acknowledgements}

\end{document}